\documentclass[onecolumn, journal,twoside]{IEEEtran}
\usepackage{graphicx}
\usepackage[T1]{fontenc}
\usepackage[bookmarks=false]{hyperref}
\usepackage[cmex10]{amsmath}
\usepackage{amssymb,amsthm}
\usepackage{thmtools}
\usepackage{cite}
\usepackage{textcomp}
\usepackage{siunitx}
\usepackage{color}
\usepackage{enumitem}
\usepackage{subfigure}
\usepackage{cases}
\usepackage{algorithm}
\usepackage{algorithmicx}
\usepackage{algpseudocode}
\usepackage{multirow}
\usepackage[font={small}]{caption}
\usepackage{bbm}
\usepackage{tcolorbox}

\makeatletter
\def\thm@space@setup{\thm@preskip=2pt
\thm@postskip=2pt \itshape}
\makeatother
\newtheoremstyle{newstyle}      
{} %Aboveskip 
{} %Below skip
{\mdseries} %Body font e.g.\mdseries,\bfseries,\scshape,\itshape
{} %Indent
{\bfseries} %Head font e.g.\bfseries,\scshape,\itshape
{.} %Punctuation afer theorem header
{ } %Space after theorem header
{} %Heading

\theoremstyle{newstyle}

\newtheorem{theorem}{Theorem}
\newtheorem{lemma}{Lemma}

\theoremstyle{definition}

\newtheorem{definition}{Definition}
\newtheorem{problem}{Problem}

\theoremstyle{remark}
\newtheorem{remark}{Remark}
\newtheorem{claim}{Claim}

\setlist[description]{style=multiline}

\let\emptyset\varnothing

\IEEEoverridecommandlockouts

\begin{document}
\sloppy

\setlength{\abovedisplayskip}{0.5mm}
\setlength{\belowdisplayskip}{0.5mm}
\setlength{\abovecaptionskip}{0.5mm}
\setlength{\belowcaptionskip}{-6pt}

%% Paper Title
%% You can use linebreaks \\ within to get better formatting as
%% desired. 
\title{How to Optimally Allocate Resources for Coded Distributed Computing?} 

%% Author names and affiliations:
%%
%% Avoiding spaces at the end of the author lines is not a problem with
%% conference papers because we don't use \thanks or \IEEEmembership.
%%
%% For several authors with only one affiliation:
%%
% \author{
%   \IEEEauthorblockN{Hui-Ting Chang and Stefan M.~Moser}
%   \IEEEauthorblockA{Department of Electrical and Computer Engineering\\
%     National Chiao Tung University (NCTU)\\
%     Hsinchu, Taiwan\\
%     Email: \{email-of-hui-ting,email-of-stefan\}@ieee.org} 
% }
%%
%% For up to three affiliations:
%%

%\author{
%  \IEEEauthorblockN{Songze~Li \thanks{This work is in part supported by NSF grants CAREER 1408639, CCF-1408755, NETS-1419632, EARS-1411244, ONR award N000141310094, a research gift from Intel, and an Okawa Foundation Research Grant.}
%}
%  \IEEEauthorblockA{Department of Electrical Engineering\\
%    University of Southern California\\
%    songzeli@usc.edu
%    } 
%  \and
%  \IEEEauthorblockN{Mohammad~Ali~Maddah-Ali}
%  \IEEEauthorblockA{Bell Labs\\
%    Alcatel-Lucent\\
%   mohammadali.maddah-ali@alcatel-lucent.com
%    }
%  \and
%  \IEEEauthorblockN{A.~Salman~Avestimehr}
%  \IEEEauthorblockA{Department of Electrical Engineering\\ 
%    University of Southern California\\
%   avestimehr@ee.usc.edu
%    }
%}

\author{Qian~Yu$^{*}$, Songze~Li$^{*}$, Mohammad~Ali~Maddah-Ali $^{\dagger}$, and A.~Salman~Avestimehr$^{*}$\\
$^{*}$ Department of Electrical Engineering, University of Southern California, Los Angeles, CA, USA \\ 
$^{\dagger}$ Nokia Bell Labs, Holmdel, NJ, USA\\
}

%
%\author{
%\IEEEauthorblockN{Songze Li, David T.H. Kao, and A. Salman Avestimehr %\thanks{This research was funded in part by one or all of these grants: %ONR N00014-09-1-0700, CCF-0917343, CCF-1117896, CNS-1213128, AFOSR %FA9550-12-1-0215, and DOT CA-26-7084-00.}
%}
%\IEEEauthorblockA{Ming Hsieh Dep. of Electrical Eng.\\
%    University of Southern California, CA, USA\\
%    Email: \{songzeli, kaod\}@usc.edu, avestimehr@ee.usc.edu}
%}

%%
%% For over three affiliations, or if they all won't fit within the width
%% of the page, use this alternative format:
%%
% \author{
%   \IEEEauthorblockN{
%     Michael Shell\IEEEauthorrefmark{1},
%     Homer Simpson\IEEEauthorrefmark{2},
%     James Kirk\IEEEauthorrefmark{3}, 
%     Montgomery Scott\IEEEauthorrefmark{3} and
%     Eldon Tyrell\IEEEauthorrefmark{4}}
%   \IEEEauthorblockA{
%     \IEEEauthorrefmark{1}School of Electrical and Computer Engineering\\
%     Georgia Institute of Technology, Atlanta, Georgia 30332--0250\\ 
%     Email: see http://www.michaelshell.org/contact.html}
%   \IEEEauthorblockA{
%     \IEEEauthorrefmark{2}Twentieth Century Fox, Springfield, USA\\
%     Email: homer@thesimpsons.com}
%   \IEEEauthorblockA{
%     \IEEEauthorrefmark{3}Starfleet Academy, San Francisco, California 96678-2391\\
%     Telephone: (800) 555--1212, Fax: (888) 555--1212}
%   \IEEEauthorblockA{
%     \IEEEauthorrefmark{4}Tyrell Inc., 123 Replicant Street, Los Angeles, California 90210--4321}
% }

\maketitle

\begin{abstract} 

Today's data centers have an abundance of computing resources, hosting server clusters consisting of as many as tens or hundreds of thousands of machines. To execute a complex computing task over a data center, it is natural to distribute computations across many nodes to take advantage of parallel processing. However, as we allocate more and more computing resources to a computation task and further distribute the computations, large amounts of (partially) computed data must be moved between consecutive stages of computation tasks among the nodes, hence the communication load can become the bottleneck. In this paper, we study the optimal allocation of computing resources in distributed computing, in order to minimize the total execution time in distributed computing accounting for both the duration of computation and communication phases. In particular, we consider a general MapReduce-type distributed computing framework, in which the computation is decomposed into three stages: \emph{Map}, \emph{Shuffle}, and \emph{Reduce}. We focus on a recently proposed \emph{Coded Distributed Computing} approach for MapReduce and study the optimal allocation of computing resources in this framework. For all values of problem parameters, we characterize the optimal number of servers that should be used for distributed processing, provide the optimal placements of the Map and Reduce tasks, and propose an optimal coded data shuffling scheme, in order to minimize the total execution time. To prove the optimality of the proposed scheme, we first derive a matching information-theoretic converse on the execution time, then we prove that among all possible resource allocation schemes that achieve the minimum execution time, our proposed scheme uses the exactly minimum possible number of servers.

\end{abstract}

\section{Introduction}

% Recently, a class of distributed computing problems was considered, where the computing task can be decomposed into two steps: "Map" and "Reduce". In \cite{DBLP:journals/corr/LiMA15, 2016arXiv160407086L}, a simple information theoretic framework was introduced, and it was shown that by applying coded multicasting in the shuffling phase, a significant gain that scales with the size of the network can be achieved. Several extensions of this result have been studies since then, including wireless distributed computing \cite{li2016scalable, globedcd16}, and distributed computing with straggling servers \cite{li2016unified}.

 In recent years, distributed systems like Apache Spark~\cite{zaharia2010spark} and computational primitives like MapReduce~\cite{dean2004mapreduce}, Dryad~\cite{Dryad}, and CIEL~\cite{Ciel} have gained significant traction, as they enable the execution of production-scale computation tasks on data sizes of the order of tens of terabytes and more. The design of these modern distributed computing platforms  is driven by  \emph{scaling out} computations across clusters consisting of as many as tens or hundreds of thousands of machines. As a result, there is an abundance of computing resources that can be utilized for distributed processing of computation tasks. However, as we allocate more and more computing resources to a computation task and further distribute the computations, a large amount of (partially) computed data must be moved between consecutive stages of computation tasks among the nodes, hence the communication load can become the bottleneck. This gives rise to an important problem: \begin{itemize}
 \item \textit{How should we optimally allocate computing resources for distributed processing of a computation task in order to minimize its total execution time (accounting for both the duration of computation and communication phases)?}
 \end{itemize}
%This problem has indeed attracted a lot of attention in recent years, and it has been broadly studied in various settings. For example, in the monitoring-based resource allocation systems (see e.g.,\cite{corsava2003intelligent,1540441}), according to a pre-defined policy, the performance metrics of the services are constantly monitored at computation resources. The resource scheduler is notified once some performance threshold is violated and then it responses by dynamically re-allocating the computing resources. On the other hand, in the estimation-based systems (see e.g., \cite{verma2011aria,lee2011heterogeneity}), upon receiving a new workload, the scheduler fetches the related job profile from the database, and estimates and assigns the required computation resources (e.g., number of servers and number of task slots) to execute this workload. When faced with multiple concurrent jobs, the resource allocation is performed considering the priorities of the jobs, the specified job deadlines and the overall throughput of the cluster. 

This problem has indeed attracted a lot of attention in recent years, and it has been broadly studied in various settings (see, e.g., \cite{corsava2003intelligent,1540441,verma2011aria,lee2011heterogeneity,kao2015hermes}). In this paper, we study resource allocation problem in the context of a recently proposed coding framework for distributed computing, namely \emph{Coded Distributed Computing}~\cite{2016arXiv160407086L}, which allows to optimally trade \emph{computation load} with \emph{communication load} in distributed computing. The key advantage of this framework is that it quantitatively captures the relation between computation time and communication time in distributed computing, which is crucial for resource allocation problems. 

More formally, we consider a general  MapReduce-type framework for distributed computing (see, e.g.,~\cite{dean2004mapreduce,zaharia2010spark}), in which the overall computation is decomposed to three stages, \emph{Map}, \emph{Shuffle}, and \emph{Reduce} that are executed distributedly across several computing nodes. In the Map phase, each input file is processed locally, in one (or more) of the nodes, to generate intermediate values. In the Shuffle phase, for every output function to be calculated, all intermediate values corresponding to that function are transferred to one of the nodes for reduction. Finally, in the Reduce phase all intermediate values of a function are reduced to the final result.

In Coded Distributed Computing, we allow redundant execution of Map tasks at the nodes, since it can result in significant reductions in data shuffling load by enabling in-network coding. In fact, in~\cite{DBLP:journals/corr/LiMA15, 2016arXiv160407086L} it has been shown that by assigning the
computation of each Map task at $r$ carefully chosen nodes, we can enable novel coding
opportunities  that reduce the communication load by exactly a multiplicative factor of the computation load $r$. For example, the communication load can be reduced by more than 50\% when each Map task is computed at only one other node (i.e., $r=2$).

Based on this framework, we consider two types of implementations: 1) \emph{Sequential  Implementation.} The above three phases take place one after another sequentially. In this case, the overall execution time $T_{\textup{sequential}} = T_{\textup{map}} + T_{\textup{shuffle}} + T_{\textup{reduce}}$. 2) \emph{Parallel Implementation.} The Shuffle phase happens in parallel with the Map phase. In this case, the overall execution time becomes $T_{\textup{parallel}} = \max \{T_{\textup{map}},T_{\textup{shuffle}}\}+T_{\textup{reduce}}$. Then the considered resource allocation problem for e.g., the sequential implementation can (informally) be formulated as the following optimization problem.

\begin{tcolorbox}
\begin{align}
\underset{\left\{\begin{subarray}{c}
  \vspace{1mm}
  \textup{Number of utilized servers}  \\
  \vspace{0.5mm}
  \textup{Placements of Map/Reduce tasks}\\
  \textup{Data shuffling scheme}
  \end{subarray}\right\}}{\min} T_{\textup{sequential}} = T_{\textup{map}} + T_{\textup{shuffle}} + T_{\textup{reduce}}
\end{align}
\end{tcolorbox}

% where the above minimization problems are over all resource allocation strategies that include the following four elements:
% \begin{enumerate}
% \item Number of servers used for computation.
% \item The placement of the Map tasks across the selected servers.
% \item Data shuffling scheme.
% \item The placement of the Reduce tasks across the selected servers.
% \end{enumerate}

In this paper, we \emph{exactly} solve the above optimization problem and its counterpart for the parallel implementation. In particular, for each implementation, we propose an optimal resource allocation scheme that exactly achieves the minimum execution time. In the proposed scheme to compute $Q$ output functions, for some design parameter $r^*$, we use a number of $Q+\lceil\frac{Q}{r^*}\rceil$ server nodes for computation. These servers are split into two groups that are termed as the ``solvers'' and the ``helpers''. There are $Q$ solver nodes, each computing a distinct Reduce function. The remaining $\lceil\frac{Q}{r^*}\rceil$ nodes are helpers, on which Map functions are computed to facilitate a more efficient data shuffling process. No Reduce function is computed on helpers themselves. In the Map phase, each input file is repetitively mapped on $r^*$ solver nodes according to a specified pattern. On the other hand, on the helper nodes, all input files are evenly partitioned and assigned for mapping, without any repetition. Then in the Shuffle phase, the communication is solely from the helpers to the solvers. In particular, based on the locally computed intermediate values in the Map phase, each helper node constructs \emph{coded multicast} messages that are simultaneously delivering required intermediate values to $r^*+1$ solvers. From these multicast messages, each solver node can decode the required intermediate values for reduction, using locally computed Map results. Finally, each solver node computes the assigned Reduce functions (hence the final output functions) locally, using the locally computed Map results and the intermediate values decoded from the messages received from the helpers.

We also prove the exact optimality of our proposed resource allocation strategies for both sequential and parallel implementations. To do that, we first derive a lower bound on the data shuffling time using any placements of the Map and the Reduce tasks. Then from this lower bound, we derive a lower bound on the minimum total execution time, and show that it is no shorter than the time achieved by the proposed strategy. At the same time, we also prove that the proposed strategy always uses exactly the minimum required number of servers to achieve the exact minimum execution time, by showing that the derived lower bound on the minimum execution time cannot be achieved with less number of servers.

~

%The idea of injecting computation redundancy to provide robustness to stagglers has also been propoposed in \cite{} [][].  the unified coding that injects redundancy for both xx and xx was in []. In this work, we do not focus on the stra effect. In all these works

\noindent \textbf{Related Work.}
The idea of injecting structured redundancy in computation to provide the coding opportunity that significantly reduces the communication load has been studied in \cite{DBLP:journals/corr/LiMA15, 2016arXiv160407086L,  globedcd16, li2016scalable, li2016unified}.  In all these works, it was assumed that the computation is carried out with a fixed number of computing nodes. Furthermore, it assumed a balanced design of the computation scheme, where the reduce jobs in the considered MapReduce-type framework have to be evenly distributed on all the nodes. Under these assumptions, they focused on characterizing the optimal tradeoff between the computation load in the Map phase, and the communication load in the Shuffle phase, by designing only the Map phase and the Shuffle phase.
In this paper, we generalize the prior works by allowing the flexibility of using an arbitrary number of servers, and unbalanced reduce task assignments on the computing nodes. We design all three phases (Map, Shuffle, and Reduce) and aim to minimize the total execution time. We also aim to minimize the usage of computing resources (nodes) while achieving the optimal performance.
In another line of research, \cite{lee2016speeding} showed that injecting redundancy in computation also provides robustness to handle straggling effects, and \cite{li2016unified} proposed a framework that takes both the straggling effect and the bandwidth usage into account. In this work, we do not focus on the straggling effect and we consider the simple model where all the nodes are computing with the same speed.

%Our main contribution is that we provide a novel and optimal design of the computation scheme, which requires assigning the computation jobs asymmetrically to the nodes, and we prove that the scheme, which only requires using a limited number of computation nodes in most cases (even though we assume that the total number of available nodes is large), uses exactly the minimum number of nodes among all computing schemes that achieve this minimum execution time.

~

The rest of the paper is organized as follows. Section \ref{sec:def} formally establishes the system model and defines the problems. Section \ref{sec:main} summarizes and discusses the main results of this paper. Section \ref{sec:scheme} describes the proposed resource allocation schemes for both sequential and parallel implementations. Section \ref{sec:conv} proves the exact optimality of the proposed schemes through matching information-theoretic converses. Section \ref{sec:conc} concludes the paper.

\section{Problem Formulation} \label{sec:def}
We consider a problem of computing $Q$ output functions from $N$ input files, for some system parameters $Q,N\in \mathbb{N}$. More specifically, given $N$ input files $w_1,\ldots,w_N \in \mathbb{F}_{2^F}$, for some $F \in \mathbb{N}$, the goal is to compute $Q$ output functions $\phi_1,\ldots,\phi_Q$, where $\phi_q:(\mathbb{F}_{2^F})^N \rightarrow \mathbb{F}_{2^B}$, $q \in \{1,\ldots,Q\}$, maps all input files to a $B$-bit output value $u_q = \phi_q(w_1,\ldots,w_N) \in \mathbb{F}_{2^B}$, for some $B \in \mathbb{N}$. 

\begin{figure}[htbp]
%\vspace{-5mm}
   \centering
   \includegraphics[width=0.4\textwidth]{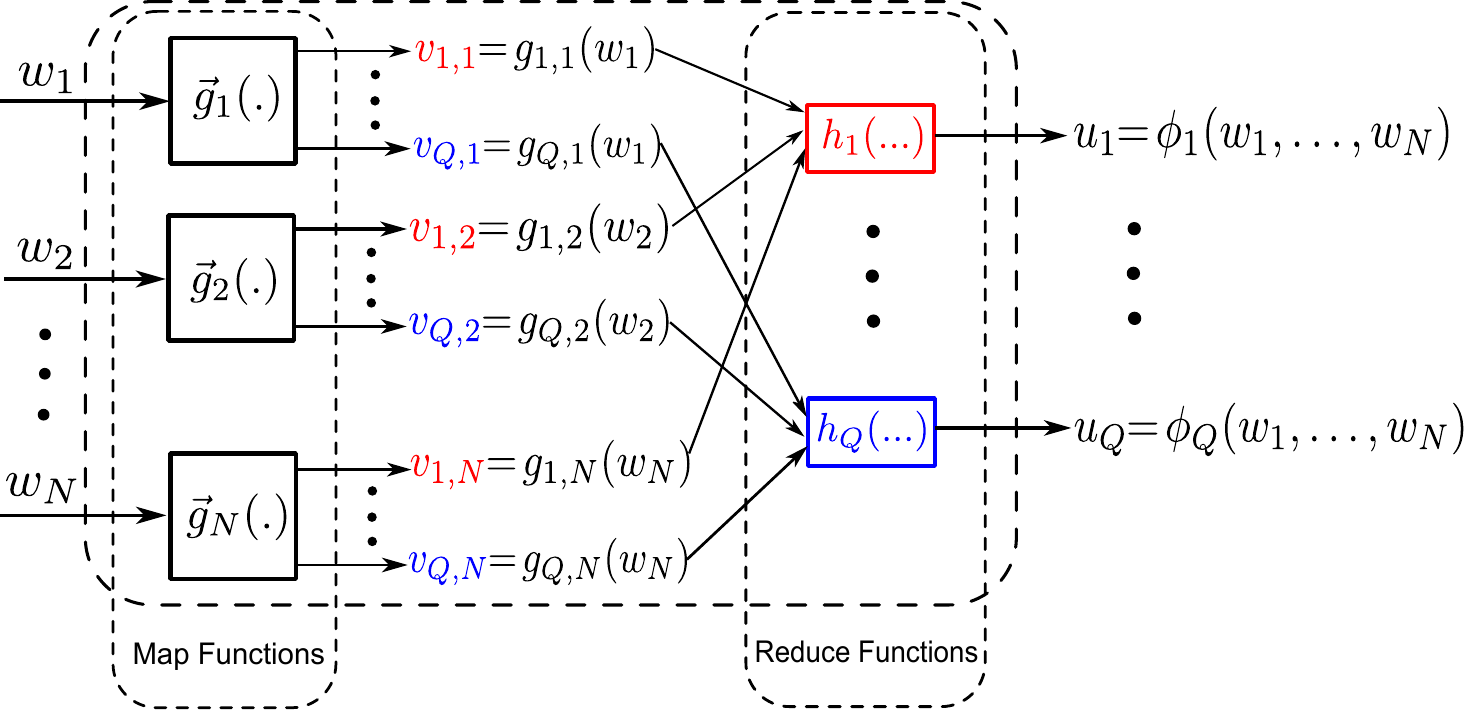}
   \caption{Illustration of a two-stage distributed computing framework. The overall computation is decomposed into computing a set of Map and Reduce functions.}
   %\vspace{-3mm}
   \label{fig:frame}
\end{figure}

We employ a MapReduce-type distributed computing structure and decompose the computation of the output function $\phi_q$, $q \in \{1,\ldots,Q\}$, as follows:
\begin{equation}\label{eq:decom}
\phi_q(w_1,\ldots,w_N) = h_q(g_{q,1}(w_1),\ldots,g_{q,N}(w_N)),
\end{equation}
where as illustrated in Fig.~\ref{fig:frame},
\begin{itemize}[leftmargin=4mm]
\item The ``Map'' functions $\vec{g}_{n} \!=\! (g_{1,n},\ldots,g_{Q,n})\!:\mathbb{F}_{2^F} \!\rightarrow \! (\mathbb{F}_{2^T})^Q$, $n \in \{1,\ldots,N\}$, maps the input file $w_n$ into $Q$ length-$T$ \emph{intermediate values} $v_{q,n}=g_{q,n}(w_n)\in  \mathbb{F}_{2^T}$, $q\in \{1,\ldots,Q\}$, for some $T \in \mathbb{N}$.
\item The ``Reduce'' functions $h_{q}: (\mathbb{F}_{2^T})^N \!\rightarrow \! \mathbb{F}_{2^B}$, $q\in \{1,\ldots,Q\}$, maps the intermediate values of the output function $\phi_q$ in all input files into the output value $u_q=h_q(v_{q,1},\ldots,v_{q,N})=\phi_q(w_1,\ldots,w_N)$. 
\end{itemize}

We perform the above computation using $K$ distributed computing servers, labelled by Server~$1,$ $\ldots,$ Server~$K$. Here the number of servers $K$ is a design parameter and can be an arbitrary positive integer. The chosen $K$ servers carry out the computation in three phases: \emph{Map}, \emph{Shuffle} and \emph{Reduce}. 

\noindent {\bf Map Phase.} 
In the Map phase, each server maps a subset of input files. For each $k \in \{1,\ldots,K\}$, we denote the indices of the files mapped by Server~$k$ as ${\cal M}_k$, which is a design parameter. Each file is mapped by at least one server, i.e., $\underset{k=1,\ldots,K}{\cup} \mathcal{M}_k=\{1,\ldots,N\}$. For each $n$ in $\mathcal{M}_k$, Server~$k$ computes the Map function $\vec{g}_n(w_n)\!=\!(v_{1,n},\ldots,v_{Q,n})$. 

\begin{definition}[Peak Computation Load]
We define the \emph{peak computation load}, denoted by $p$, $0 \leq p \leq 1$, as the maximum number of files mapped at one server, normalized by the number of files $N$, i.e., $p \triangleq \frac{\max\limits_{k=1,\ldots,K} |{\cal M}_k|}{N}$. $\hfill \Diamond$
\end{definition}
\vspace{-0.5mm}

We assume that all servers are homogeneous and have the same processing capacity. The average time a server spends in the Map phase is linearly proportional to the number of Map functions it computes, i.e., the average time for a server to compute $n$ Map functions is $c_{\textup{m}} \frac{n}{N}$, for some constant $c_{\textup{m}} >0$.  Also, since the servers compute their assigned Map functions simultaneously in parallel, we define the \emph{Map time}, denoted by $T_{\textup{map}}$, as the average time for the server mapping the most files to finish its computations, i.e., 
\begin{align}
T_{\textup{map}} =  \max \limits_{k=1,\ldots,K} c_{\textup{m}}\tfrac{|{\cal M}_k|}{N} =c_{\textup{m}} p.
\end{align}

The minimum possible Map time can be arbitrarily close to $0$, assuming $N$ is large. This minimum Map time can be achieved by using a large number of servers, and letting the $N$ Map tasks be uniformly assigned to these servers without repetition.
%To minimize the Map time, we need at least $N$ servers, each computing a unique Map function. In this way we can achieve the minimum Map time of $c_{\textup{m}}/N$. %TODO

\noindent {\bf Shuffle Phase.}
We assign the tasks of computing the $Q$ output functions across the $K$ servers, and denote the indices of the output functions computed by Server~$k$, $k=1,\ldots,K$, as $\mathcal{W}_k$, which is also a design parameter. Each output function is computed exactly once at some server, i.e., 1) $\underset{k=1,\ldots,K}{\cup} {\cal W}_k = \{1,\ldots,Q\}$, and 2)  ${\cal W}_j \cap {\cal W}_k = \emptyset$ for $j \neq k$. 

To compute the output value $u_q$ for some $q \in \mathcal{W}_k$, Server~$k$ needs the intermediate values that are \emph{not} computed \emph{locally} in the Map phase, i.e., $\{v_{q,n}: q \in \mathcal{W}_k, n \notin \mathcal{M}_k\}$. After the Map phase, the $K$ server proceed to exchange the needed intermediate values for reduction. We formally define a \emph{shuffling scheme} as follows:
\begin{itemize}[leftmargin=4mm]
\item Each server~$k$, $k \in \{1,\ldots,K\}$, creates a message $X_k$ as a function of the intermediate values computed locally in the Map phase, i.e., $X_k = \psi_k\left(\{\vec{g}_n:n \in \mathcal{M}_k\}\right)$, and multicasts it to a subset of $1\leq j \leq K-1$ nodes. 
\end{itemize}

\begin{definition}[Communication Load]
We define the \emph{communication load}, denoted by $L$, $0 \leq L \leq 1$, as the total number of bits communicated by all server in the Shuffle phase, normalized by $QNT$ (which equals the total number of bits in all intermediate values $\{v_{q,n}: q\in \{1,\ldots,Q\}, n \in \{1,\ldots,N\} \}$).\footnote{In this paper, we assume that the cost of multicasting to multiple servers is the same as unicasting to one server.}  $\hfill\Diamond$
\end{definition}

For some constant $c_\textup{s} >0$, we denote the bandwidth of the shared link connecting the servers as $1/c_\textup{s}$. Thus given a communication load of $L$, the \emph{Shuffle time}, denoted by $T_{\textup{shuffle}}$, is defined as
\begin{align}
T_{\textup{shuffle}} = c_\textup{s} L.
\end{align}

The minimum possible Shuffle time is $0$. It can be achieved by having each of the servers assigned to compute the Reduce functions map all $N$ files locally.

\noindent {\bf Reduce Phase.} 
Server $k$, $k\in\{1,\ldots,K\}$, uses the local Map results $\{\vec{g}_n: w_n\in \mathcal{M}_k\}$ and the received messages $X_1,\ldots,X_K$ in the Shuffle phase to construct the inputs to the assigned Reduce functions in $\mathcal{W}_k$, and computes the output value $u_q=h_q(v_{q,1}\ldots v_{q,N})$ for all $q \in \mathcal{W}_k$. 

Similar to the computations of the Map functions, the average time for a server to compute $q$ Reduce functions is $c_\textup{r} q$, for some constant $c_\textup{r} >0$. The servers compute their assigned Reduce functions simultaneously in parallel. We define the \emph{Reduce time}, denoted by $T_{\textup{reduce}}$, as the average time for the server reducing the most output functions to finish its computations, i.e., 
\begin{align}
T_{\textup{reduce}} = c_\textup{r} \underset{k=1,\ldots,K}{\max} |{\cal W}_k|.
\end{align}

The minimum Reduce time equals $c_{\textup{r}}$. To minimize the Reduce time, we need at least $Q$ servers, and each computing a unique Reduce function. %In this way we can achieve the minimum Reduce time of $c_\textup{r}$.  %TODO:

In this setting, we are interested in designing distributed computing schemes, which includes the selection of $K$, the assignment of the Map tasks $\boldsymbol{{\cal M}}\triangleq({\cal M}_1\ldots,{\cal M}_K)$, the assignment of the Reduce tasks $\boldsymbol{{\cal W}}\triangleq({\cal W}_1\ldots,{\cal W}_K)$, and the design of the data shuffling scheme, in order to minimize the overall execution time to accomplish the distributed computing tasks.

Specifically, the overall execution time is the total amount of time spent executing the above three phases of the computation. In this paper, we consider the following two types of implementations.
\begin{enumerate}[leftmargin=5mm]
\item \emph{Sequential Implementation.} For the sequential implementation, the three phases take place one after another sequentially, e.g., the Shuffle phase does not start until all servers have completed their Map computations. In this case, the overall execution time $T_{\textup{sequential}} = T_{\textup{map}} + T_{\textup{shuffle}} + T_{\textup{reduce}}$.
\item \emph{Parallel Implementation.} For the parallel implementation, the Shuffle phase happens in parallel with the Map phase, i.e., a server communicates a message as soon as the intermediate values needed to construct the message is calculated locally from the Map functions. In this case, the overall execution time becomes $T_{\textup{parallel}} = \max \{T_{\textup{map}},T_{\textup{shuffle}}\}+T_{\textup{reduce}}$.
\end{enumerate} 

%TODO: add definition
% For a K, define minimum T_K^*, and T
% def K^*
To design the optimal distributed computing scheme that minimizes the execution time while using as few servers as possible, we need to answer the following questions:

\begin{itemize}
\item What is the minimum possible execution time?
\item What is the minimum number of servers needed to achieve the minimum possible execution time?
\item How to place the Map, Reduce tasks and design the data shuffling scheme to achieve the minimum execution time?
\end{itemize}

To answer these questions, we formulate them into the following problem: 

\begin{problem}[Optimal Resource Allocation] %example environment

Consider a computing task with parameters $Q$ and $N$. Given a certain number of servers $K$, a Map task assignment $\boldsymbol{{\cal M}}$ and a Reduce task assignment $\boldsymbol{{\cal W}}$ on these servers, we say a shuffling scheme is \emph{valid} if, for any possible outcomes of the intermediate values $v_{q,n}$, each server can decode all its needed intermediate values based on the values that are locally computed in the map phase and the messages received during the shuffle phase.

Suppose we always use valid shuffling schemes with minimum shuffling time. We denote the resulting execution times given $K$, $\boldsymbol{{\cal M}}$ and $\boldsymbol{{\cal W}}$ by $T^*_{\textup{sequential}}(K, \boldsymbol{{\cal M}},\boldsymbol{{\cal W}})$ and $ T^*_{\textup{parallel}}(K, \boldsymbol{{\cal M}},\boldsymbol{{\cal W}})$. 
%Fixing the number of participating servers $K$, we define $K$\emph{-server minimum execution times}, denoted by $T^*_{\textup{sequential}}(K)$ and $ T^*_{\textup{parallel}}(K)$, to be the minimum execution time given the optimal Map and Reduce task assignment, for large $N$. Rigorously,
%\begin{align}
%T^*_{\textup{sequential}}(K)&=\limsup_{N\rightarrow\infty} \min_{\boldsymbol{{\cal M}},\boldsymbol{{\cal W}}}{T^*_{\textup{sequential}}(\boldsymbol{{\cal M}},\boldsymbol{{\cal W}})},\\
%T^*_{\textup{parallel}}(K)&=\limsup_{N\rightarrow\infty} \min_{\boldsymbol{{\cal M}},\boldsymbol{{\cal W}}}{T^*_{\textup{parallel}}(\boldsymbol{{\cal M}},\boldsymbol{{\cal W}})}.
%\end{align}
Assuming $N$ is large, we aim to find the \emph{minimum execution times} over all possible designs, which can be rigorously defined as follows:
%We aim to find the \emph{minimum execution times}, which are defined as follows:
\begin{align}
T^*_{\textup{sequential}}&=\inf_{K,\boldsymbol{{\cal M}},\boldsymbol{{\cal W}}}{T^*_{\textup{sequential}}(K,\boldsymbol{{\cal M}},\boldsymbol{{\cal W}})},\\
T^*_{\textup{parallel}}&=\inf_{K,\boldsymbol{{\cal M}},\boldsymbol{{\cal W}}}{T^*_{\textup{parallel}}(K,\boldsymbol{{\cal M}},\boldsymbol{{\cal W}})}.
\end{align}

We are also interested in finding the minimum number of servers required to exactly achieve the minimum execution time for large $N$, denoted by $K^*_{\textup{sequential}}$ and $K^*_{\textup{parallel}}$, defined as follows
\begin{align}
K^*_{\textup{sequential}}&=\min\{K\in\mathbb{N}\ |\    \min_{\boldsymbol{{\cal M}},\boldsymbol{{\cal W}}} T^*_{\textup{sequential}}(K,\boldsymbol{{\cal M}},\boldsymbol{{\cal W}})=T^*_{\textup{sequential}}\},\\
K^*_{\textup{parallel}}&=\min\{K\in\mathbb{N}\ |\   \min_{\boldsymbol{{\cal M}},\boldsymbol{{\cal W}}}T^*_{\textup{parallel}}(K,\boldsymbol{{\cal M}},\boldsymbol{{\cal W}})=T^*_{\textup{parallel}}\}.
\end{align}
If the minimum in any of the above equations does not exist, we say the corresponding $T^*_{\textup{sequential}}$ or $T^*_{\textup{parallel}}$ can not be \emph{achieved using finite number of servers}.

%Because the minimum shuffling time for valid shuffling schemes is completely determined by $\boldsymbol{{\cal M}}$ and $\boldsymbol{{\cal W}}$, 
Besides, we want to find the optimal computing schemes that minimizes the execution time while using the minimum number of servers. Specifically, for each implementation, we want to construct a Map task assignment $\boldsymbol{\mathcal{M}}$, a reduce task assignment $\boldsymbol{\mathcal{W}}$, and a valid shuffling scheme design, that achieve the minimum execution time using the minimum number of servers. $\hfill \Diamond$   \label{prob}
\end{problem}

In this paper, we answer all the questions mentioned in the above problem. Interestingly, some of the answers match the intuition and some do not. For example, the coding gain in our proposed optimal scheme is obtained through coded multicasting, which agrees with the intuition. However, counter intuitively, the optimal scheme requires a non-symmetric design, where the servers are classified into two groups. One group is only assigned Map and Reduce tasks, focusing on computing the output functions; while the other group only does Map and Shuffle, focusing on delivering the intermediate results and exploiting the multicast opportunity. Also, the intuition may suggest that by using more servers, we may always be able to further reduce the execution time. However, we show that in most cases, the minimum execution time can be exactly achieved using finitely many servers, and the minimum execution time can not be further reduced after the number of servers passes a threshold.

%For example, in the optimal computing scheme, 
% (Intuitively map (and shuffle) should be uniform, but here is not) (finite K)

\section{Main Results}\label{sec:main}
For the sequential implementation, we characterize the minimum execution time $T_{\textup{sequential}}^*$ and the minimum number of servers to achieve $T_{\textup{sequential}}^*$ in the following theorem.

\begin{theorem}[Sequential Implementation]\label{th:seq}
	For a distributed computing application that computes $Q$ output functions, $T_{\textup{sequential}}^*$ defined in Problem \ref{prob} is given by 
	\begin{align}
		T_{\textup{sequential}}^*&= c_{\textup{m}} \tfrac{r^*}{Q} + c_\textup{s} \tfrac{Q-r^*}{Q(r^*+1)}+c_\textup{r},\label{eq:th1}
	\end{align}
	 where $r^*$ is defined as follows:
% and parallel implementations in the following theorem.
%	Let $r^*$ denotes the largest parameter $r$ that minimizes the execution time in (\ref{eq:th1}), i.e.,
	\begin{align}\label{rstars}
	r^*=\max \underset{r \in \{0,1\ldots,Q\}}{\textup{argmin}} \ c_\textup{m} \frac{r}{Q} + c_\textup{s} \frac{Q-r}{Q(r+1)}.
	\end{align}
	We can show that the above execution time can be exactly achieved using a finite number of servers if and only if $r^*\neq 0$.  For $r^*\neq 0$, $K_{\textup{sequential}}^*$ defined in Problem \ref{prob} is given by
		\begin{align}
		K_{\textup{sequential}}^* = \begin{cases} Q+\lceil\tfrac{Q}{r^*}\rceil, & 0< r^* < Q,\\
		Q, & r^* = Q.
		\end{cases}
		\end{align}
		%where $r^*=\underset{r \in \{0,1\ldots,Q\}}{\textup{argmin}} \ c_m \frac{r}{Q} + c_s \frac{1-\frac{r}{Q}}{r+1}$.
	\end{theorem}
	%Amdahl's law
	
	%In prior works, it is assumed that given a fixed total number of reduce jobs is fixed, the jobs are uniformly distributed across the servers. In this paper, we allow the flexibility to arbitrarity assign the reduce jobs to the servers, as long as all jobs are handled. arbitrary num of server, and 

		\begin{remark}
	    The above theorem generalizes the prior works on coded distributed computing, \cite{DBLP:journals/corr/LiMA15,2016arXiv160407086L,globedcd16, li2016scalable}, by allowing the flexibility of using arbitrary number of servers and arbitrary reduce task assignments on the servers. 
	    	 In prior works, it is assumed that all the $Q$ Reduced tasks are uniformly assigned to all the servers. %Consequently, in the priorly proposed scheme, all servers have jobs in all three phases: map and shuffle, and reduce. 
	    	 In this paper, we will see that by focusing on the execution time and allowing using arbitrary number of servers, the optimal scheme naturally requires a certain Reduce task assignment, where each server either reduce $1$ function, or does not reduce at all. %In most cases, this optimal reduce job assignment is not uniform. 
	    	%Besides, the optimal scheme requires that all shuffling jobs are handeled by servers that are not assigned reduce jobs. 
	  %thus we no longer require any assumption on the key distribution.
	 %In the optimal solution, there are two classes of servers, some server executes both map and reduce, and some server only do mapping but not reducing at all. 
	 To simplify the discussion, we refer to the servers that are assigned Reducing tasks as \emph{solvers}, and we refer to the rest of the servers %(i.e., the servers that only do mapping and shuffling jobs) 
	 as \emph{helpers}. %TODO:(each server do 1 or 0)
%	 (In prior, all do map reduce, and unif)
	    %In prior works, it is assumed that the number of jobs at each server is fixed, and more over, the jobs are uniformly distributed across the servers. In this paper, we will see that by focusing on the execution time and allowing using arbitrary number of servers, the optimal scheme naturally requires a certain reduce job assignment (the total job is still Q, but the assignment (in most cases) is not uniform), thus we no longer require any assumption on the key distribution. (In prior, all do map reduce, and unif)(the assignment is not uniform) (In opt sol, we have two categories of servers, some user do both map and reduce, some user do map but do nothing in reduce, we call group 1 and group 2) (each server do 1 or 0)
	\end{remark}
	
		\begin{remark}
		
			To achieve the above minimum execution time, we propose a distributed computing scheme, where each server maps no more than $\frac{r^*}{Q}$ fraction of the files in the database, with communication load of $\frac{Q-r^*}{Q(r^*+1)}$. In the proposed achievability scheme, we will see each file repetitively mapped on $r^*$ solvers. Having this redundancy in the Map phase has two advantages: first of all, more computation enhances the local availability of the intermediate values, thus each solver only needs values from $1-\frac{r^*}{Q}$ fraction of the files from the the shuffling phase; secondly, mapping the same file at multiple servers allows delivering intermediate values through coded multicasting, and a coding gain of $r^*+1$ is achieved in the proposed delivery scheme.
		
		\end{remark}
	
		\begin{remark}  
		Similar to the prior works \cite{DBLP:journals/corr/LiMA15,2016arXiv160407086L,globedcd16, li2016scalable}, the trade off between computation load and communication load can be established, and the above theorems  demonstrate how the optimal peak computation load can be chosen based on %the parameter of the problem and 
		the trade off. 
	\end{remark}
	
		\begin{remark}%TODO: pass message clearly, why increasing K won't help (why not 0 computation, some sommunication, refer to remark 0)
		We prove the exact optimality of the proposed scheme through a matching information theoretic converse, which is provided in section \ref{sec:conv}.
		%In Theorem \ref{th:seq}, 
		We observe that in most cases, using a finite number of servers is sufficient to exactly achieve the lower bound of the minimum execution time, which means that the execution time cannot be further reduced by using more servers than the provided $K_{\textup{sequential}}^*$. 
		%explain, also key to reduce job
		This is due to the fact that the coded multicasting opportunity, which is essential to achieving the minimum communication load, relies on mapping the files repetitively on the solvers. Because the total number of Reduce functions is fixed, the number of solvers is upper bounded by $Q$ even if we use infinitely many servers. Consequently, by using a large number of servers, reducing the peak computation load on the solvers will inevitably reduce the number of times that each file is repetitively mapped on the solvers, which consequently hurts the coded multicasting opportunity and increases the communication load. Hence, the entire benefit of using more than $Q$ servers is to reduce the computation load of the helpers, until the computation load of the solvers becomes the bottleneck. Further increasing the number of servers will not affect the computation-communication trade-off.
		
		Conversely, Theorem \ref{th:seq} also indicates that, when using fewer servers than the suggested minimum number ($K_{\textup{sequential}}^*$), the resulting computing scheme must be strictly suboptimal. This is due to the fact that only the helpers can fully utilize the coded multicasting opportunity during the shuffling phase. Hence, to achieve the minimum communication load, no shuffling job should be handled by the solvers, and we need sufficient helpers to map enough files in order to obtain enough information to support the shuffling phase, without becoming the bottleneck of the peak computation load. 
		%(pass message clearly, why not increase, add intuitive understanding)
	\end{remark}

	\begin{remark} %TODO: (more direct)
	From theorem \ref{th:seq}, we observe that the optimal solution always requires using at least $Q$ servers, which is because any computing scheme having a server reducing more that one function is strictly suboptimal (will be proved later), so at least $Q$ solvers are needed to compute all the Reduce functions. 
	
	In addition, we note that  $K^*_{\textup{sequential}}$, is a decreasing function of $r^*$, and consequently an increasing function of $\frac{c_\textup{m}}{c_\textup{s}}$, which can be explained as follows:
	%Note that when $Q$ is fixed, the minimum number of servers is a function of $r^*$, which only depends on $\frac{c_\textup{m}}{c_\textup{s}}$. 
	When $\frac{c_\textup{m}}{c_\textup{s}}$ increases, the computation time for mapping one file becomes relatively larger, therefore it is better to pick a computing scheme with larger communication load and smaller computation load. To reduce the computation load, $r^*$, the number of times each file is repetitively mapped on all the solvers, should be decreased. As a result, the peak computation load on the helpers also decreases, and thus more helpers are needed to make sure that each file  needed for the shuffling phase is mapped on at least one helper. 
	\end{remark}
	
		\begin{remark} %TODO: add intuition
		%(intuitive, one is r, other is 1/s, for opt, these approx same, so we have sqrt)
			If we ignore the integrality constraint, $r^*$ and $K_{\textup{sequential}}^*$ can be approximated as follows:
			\begin{align}
			r^*&\approx \sqrt{(Q+1)\frac{c_\textup{s}}{c_\textup{m}}}-1\approx \sqrt{Q\frac{c_\textup{s}}{c_\textup{m}}}\\
			K_{\textup{sequential}}^*&\approx Q+Q/(\sqrt{(Q+1)\frac{c_\textup{s}}{c_\textup{m}}}-1)\approx Q+\sqrt{Q\frac{c_\textup{m}}{c_\textup{s}}}.
			\end{align}
			Interestingly, $r^*$ is approximately proportional to the square root of $\frac{c_\textup{s}}{c_\textup{m}}$, while the number of helpers (i.e., $K^*_{\textup{sequential}}-Q$) is inversely proportional to the square root of $\frac{c_\textup{s}}{c_\textup{m}}$. Hence if the computation time of mapping one file is increased by $4$ times, $r^*$ should be halved, and the number of helpers should be doubled. 
			
			 We have the following explanation: In the optimal computing scheme proposed in this paper, the computation time is proportional to $c_{\textup{m}}r$, and the communication time is approximately $c_{\textup{s}}/r$, where $r$ is the number of times each file is repetitively mapped on all solvers. To minimize the total execution time, the design parameter should balance the time used in these two phases, which results that $r^*$ should be approximately proportional to the square root of $\frac{c_\textup{s}}{c_\textup{m}}$. Besides, in most cases the helpers should map all files in the database in order to execute the shuffling functions. Hence the minimum number of helpers (i.e., $K^*_{\textup{sequential}}-Q$) should be inversely proportional to the computation load, which should consequently be inversely proportional to the square root of $\frac{c_\textup{s}}{c_\textup{m}}$.  
	\end{remark}

\begin{remark} \label{rem:uncoded_seq}
    As we have discussed, achieving the minimum possible communication load relies on exploiting local availabilities and allowing coded multicasting. As a comparison, we consider computing designs where the opportunity of multicasting during the shuffling phase is not utilized, i.e., the shuffling phase is \emph{uncoded}. The minimum execution time %and the needed number of servers to achieve the minimum execution time for sequential implementation 
    is given as follows: 
	%{\bf To Do: add impact}
	\begin{align}
	T_{\textup{sequential, uncoded}}^*&=\min_{r\in \{0,Q\}} c_\textup{m} \tfrac{r}{Q} + c_\textup{s} ({1-\frac{r}{Q}})+c_\textup{r}\\
	&=\min\{c_\textup{m} , c_\textup{s} \}+c_\textup{r}.
	\end{align}
%	\begin{align}
%	T_{\textup{parallel, uncoded}}^*&=\min_{r\in [0,Q]} \max \{c_\textup{m} \tfrac{r}{Q}, c_\textup{s} ({1-\frac{r}{Q}})\}+c_\textup{r}\\
%	&=\frac{c_\textup{m}c_\textup{s}}{c_\textup{m}+c_\textup{s}}+c_\textup{r}.
%	\end{align}
%	\begin{align}
%	K_{\textup{sequential, uncoded}}^* = \max\{Q, \lceil\frac{Q}{r^*}\rceil\}.
%	\end{align}
	The above execution time can be achieved using uncoded computing scheme with finite number of servers if and only if $c_{\textup{m}}\leq c_{\textup{s}}$, and the minimum needed number of server in this case equals $Q$.
	%Similar to the prior works \cite{DBLP:journals/corr/LiMA15, 2016arXiv160407086L}, 
	
	Compared to the uncoded scheme, a large coding gain that scales with the size of the problem can be achieved by exploiting coded multicasting opportunities during the shuffling phase. For example, when $c_m=c_s$, the execution time for the Map and Shuffle phase of the optimal coded scheme grows as $\Theta(Q^{-\frac{1}{2}})$, while the execution time of the uncoded scheme remains constant.
	
	The two schemes also requires different number of servers to achieve the minimum execution time. For the uncoded computing scheme, at most $Q$ servers are needed to achieve the minimum cost, unless the computing power of $Q$ servers are not sufficient to map the entire database% (i.e., $r<1$)
	; while for the coded computing scheme, in most cases more that $Q$ servers are needed to achieve the minimum execution time.
	This is due to the fact that in the coded computing scheme, the Reduce tasks and the shuffling jobs are handled by disjoint groups of servers in order to fully maximize the coding gain, and hence extra servers  are needed to optimize the performance. However in the uncoded scheme, the only use of non-solver nodes is to provide extra computing power. Hence when $Q$ servers are sufficient to map the entire database, using more servers does not reduce the execution time.
	%extra server do not provide coding gain
\end{remark}

For the parallel implementation, we characterize the minimum execution time, and the minimum number of servers to achieve $T_{\textup{sequential}}^*$ in the following theorem

\begin{theorem}[Parallel Implementation]\label{th:par}
For a distributed computing application that computes $Q$ output functions, $T_{\textup{parallel}}^*$ defined in Problem \ref{prob} is given by 
\begin{align}
	T_{\textup{parallel}}^*&=\max\{c_\textup{m} \tfrac{r^*}{Q} , c_\textup{s}\cdot \textup{Conv}(\tfrac{Q-r^*}{Q(r^*+1)})\}+c_\textup{r},\label{eq:th3}
\end{align}
 where $\textup{Conv}(f(\cdot))$ denotes the lower convex envelope of points $\{(r,f(r))\ |\ r\in\{0,1,...,Q\}\}$, and $r^*$ is defined as follows:
	\begin{align}\label{rstarp}
	r^*= \underset{0\leq r\leq Q}{\textup{argmin}} \ \max\{c_\textup{m} \tfrac{r}{Q} , c_\textup{s}\cdot \textup{Conv}(\tfrac{Q-r}{Q(r+1)})\}.
	\end{align}
	We can show that the above execution time can be exactly achieved using a finite number of servers, and $K_{\textup{parallel}}^*$ defined in Problem \ref{prob} is given by
\begin{align}
K_{\textup{parallel}}^* = \begin{cases} Q+\lceil\tfrac{Q}{r^*}\rceil, & r^* \leq Q-1,\\
Q+\lceil\tfrac{Q(Q-r^*)}{r^*}\rceil, & r^* > Q-1.
\end{cases}
\end{align}
%where $r^*=\underset{r \in \{0,1\ldots,Q\}}{\textup{argmin}} \ c_m \frac{r}{Q} + c_s \frac{1-\frac{r}{Q}}{r+1}$.
\end{theorem}

%\begin{theorem}
%For the parallel implementation, the minimum required number of servers to achieve the minimum execution time is
%\begin{align}
%K_{\textup{parallel}}^* = \begin{cases} Q+\lceil\tfrac{Q}{r^*}\rceil, & r^* < Q,\\
%Q, & r^* \geq Q,
%\end{cases}
%\end{align}
%where $r^*=\underset{r}{\textup{argmin}} \max\{c_m \frac{r}{Q} , c_s \frac{1-\frac{r}{Q}}{r+1}\}$.
%\end{theorem}

%\begin{theorem}
%	For Sequential Implementation and Parallel Implementation, %TODO1. one for seq one for par, 2. fix par
%	\begin{align}
%	K^*= Q+\lceil\frac{Q}{r^*}\rceil
%	\end{align}
%	i f  $r^*<Q$, and
%	\begin{align}
%	K^*= Q
%	\end{align}
%	if  $r^*\geq Q$,
%	where 
%	\begin{align}
%	r^*=\underset{r}{\textup{argmin}} \ c_m \frac{r}{Q} + c_s \frac{1-\frac{r}{Q}}{r+1}
%	\end{align}
%	for Sequential Implementation, and 
%	\begin{align}
%	r^*=\max\{c_m \frac{r}{Q} , c_s \frac{1-\frac{r}{Q}}{r+1}\}
%	\end{align}
%	for Parallel Implementation
%	, assuming $N$ is large.
%\end{theorem}

%{\bf To Do: Challenge (what is new)}
%{\bf To Do: Compare with uncoded} (also uncoded does not utilize more servers)

	\begin{remark}
	    The above theorem generalized the prior works  \cite{DBLP:journals/corr/LiMA15,2016arXiv160407086L,globedcd16, li2016scalable}, by allowing the flexibility of using an arbitrary number of servers and arbitrary Reduce task assignments on the servers. Similar to the sequential implementation, the optimal scheme for parallel implementation also requires a certain Reduce task assignment, where each server either reduces $1$ function or does not reduce at all. %Besides, the optimal scheme also requires that all shuffling jobs are handled by servers that are not assigned reduce jobs. 
	 Thus, we continue to use the names \emph{solvers} and \emph{helpers} for the parallel implementation. 
	\end{remark}
	
		\begin{remark}
	To achieve the above minimum execution time, we propose a distributed computing scheme, where each server maps no more than $\frac{r^*}{Q}$ fraction of the files in the database, with communication load of $\textup{Conv}(\frac{Q-r^*}{Q(r^*+1)})$. %and in most cases, each file in the dataset is mapped by more than one server. 
	Similar to the sequential case, each file is repetitively mapped $r^*$ times. This redundancy enhances the local availability of the intermediate values, and allows delivering intermediate values through coded multicasting. Hence, by following the same argument, we can achieve the same computation-communication trade off achieved by the scheme used in sequential implementations. However, given the same computation-communication trade off, the above theorem indicates that the optimal peak computation load should be chosen differently compared to the sequential case, in order to minimum the execution time for parallel implementation.
	%To execute the reduce job of $Q$ keys, all servers needs a total of $QN$ intermediate keys, which is greater than the communication load of the optimal scheme. The reduction in the communication load comes from two part:     
	\end{remark}
	
		\begin{remark}
		We prove the exact optimality of the proposed scheme through a matching information theoretic converse, which is provided in section \ref{sec:conv}.
		%In Theorem \ref{th:seq}, 
		We note that for parallel implementation, using finite number of servers is sufficient to exactly achieve the minimum execution time. 
		Conversely, the statement in theorem \ref{th:seq} also indicates that when using less servers than the suggested minimum number ($K_{\textup{parallel}}^*$), the resulting computing scheme must be strictly suboptimal. Both statements can be understood exactly the same way as discussed for the sequential implementation.
	\end{remark}
	
	\begin{remark}
	From theorem \ref{th:par}, we observe that the optimal solution always requires using at least $Q$ servers. In addition, we note that  $K^*_{\textup{parallel}}$, is a decreasing function of $r^*$, and consequently an increasing function of $\frac{c_\textup{m}}{c_\textup{s}}$. Both observations can be understood exactly the same way as discussed for the sequential implementation.
	\end{remark}
	
		\begin{remark} %TODO: add intuition
		%(intuitive, one is r, other is 1/s, for opt, these approx same, so we have sqrt)
			If we ignore the integrality constraint, $r^*$ and $K_{\textup{parallel}}^*$ can be approximated as follows:
			\begin{align}
			r^*\approx \sqrt{Q\frac{c_\textup{s}}{c_\textup{m}}+(\frac{c_\textup{s}/c_\textup{m}+1}{2})^2}&-\frac{c_\textup{s}/c_\textup{m}+1}{2}\approx \sqrt{Q\frac{c_\textup{s}}{c_\textup{m}}}\\
			K_{\textup{parallel}}^*\approx Q+Q/r^*&\approx Q+\sqrt{Q\frac{c_\textup{m}}{c_\textup{s}}}.
			\end{align}
			Similar to the sequential case, $r^*$ is approximately proportional to the square root of $\frac{c_\textup{s}}{c_\textup{m}}$, while the number of helpers (i.e., $K^*_{\textup{parallel}}-Q$) is inversely proportional to the square root of $\frac{c_\textup{s}}{c_\textup{m}}$. Both approximations can be explained through the same arguments used for the sequential implementation.
	\end{remark}

\begin{remark}
    %We compare the execution time of the optimal design to the minimum execution time achieved 
    We consider the minimum execution time of the uncoded scheme, which is given as follows: 
	\begin{align}
	T_{\textup{parallel, uncoded}}^*&=\min_{r\in [0,Q]} \max \{c_\textup{m} \tfrac{r}{Q}, c_\textup{s} ({1-\frac{r}{Q}})\}+c_\textup{r}\\
	&=\frac{c_\textup{m}c_\textup{s}}{c_\textup{m}+c_\textup{s}}+c_\textup{r}.
	\end{align}
    The above execution time can be achieved using $K_{\textup{parallel, uncoded}}^* = \max\{Q, \lceil\frac{Q}{r^*}\rceil\}$ servers.
    
	Compared to the uncoded scheme, a large coding gain that scales with the size of the problem is achieved using the proposed coded scheme. For example, when $c_m=c_s$, the execution time for the Map and Shuffle phase of the optimal coded scheme grows as $\Theta(Q^{-\frac{1}{2}})$, while the execution time of the uncoded scheme remains constant.
	
	The two schemes also requires different number of servers to achieve the minimum execution time. For the uncoded scheme, at most $Q$ servers are needed to achieve the minimum cost, unless the computing power of $Q$ servers are not sufficient to map the entire database; while for the coded computing scheme, in most cases more that $Q$ servers are needed to achieve the minimum execution time. 
	This is due to the fact that uncoded scheme failed to exploit the coded multicast opportunity, as explained in Remark \ref{rem:uncoded_seq}.
\end{remark}

\section{Achievability Schemes}\label{sec:scheme}

%In this section, we proposed an unified distributed computing scheme that could be used to construct computing schemes that achieve the minimum costs mentioned in Section \ref{sec:main} for both sequential and parallel implementations. We also show that the constructed scheme uses minimum number of servers that is required to achieve the minimum cost. 
In this section, we construct achievability schemes that achieve the minimum execution time mentioned in Section \ref{sec:main}, using the minimum number of servers. We start by giving an illustrative example on how to build an optimal scheme for sequential implementation given a specific set of values of problem parameters. Then we proceed to present the optimal achievability scheme for general parameters. The optimal achievability schemes for the parallel implementation is described in Appendix \ref{app:par}.

%First of all, we show that for any parameters $r\in \{0,1,...,Q-1\}$, $K>Q$, we can find a coded distributed computing scheme using $K$ nodes, that $Q$ of the nodes each reducing $1$ function, each solver maps $\frac{r}{Q}$ fraction of the files, each helper maps $\frac{1}{K-Q}$ fraction of the files, and the communication load equals $\tfrac{Q-r}{Q(r+1)}$. 
%Then we demonstrate how the above unified scheme can be used to construct the optimal computing schemes for both implementations.

%First, we present an illustrative example to demonstrate the key idea of the unified scheme.

\subsection{Illustrative Example}% for $N=6$, $Q=3$, $c_{\textup{m}}=1$, $c_{\textup{s}}=4$ and $c_{\textup{r}}=1$}%, $K=5$, and $r=2$}

We present an illustrative example of the optimal achievability scheme for a given set of parameters: $N=6$, $Q=3$, $c_{\textup{m}}=1$, $c_{\textup{s}}=2$ and $c_{\textup{r}}=1$. According to Theorem \ref{th:seq}, we choose design parameter $r^*=2$ and use $K^*_{\textup{sequential}}=5$ servers. We let servers $1,2$, and $3$ reduce functions $1,2$, and $3$, respectively. %Hence the reduce time equals $$.

\begin{figure}[htbp]
	%\vspace{-5mm}
	\centering
	\includegraphics[width=0.5\textwidth]{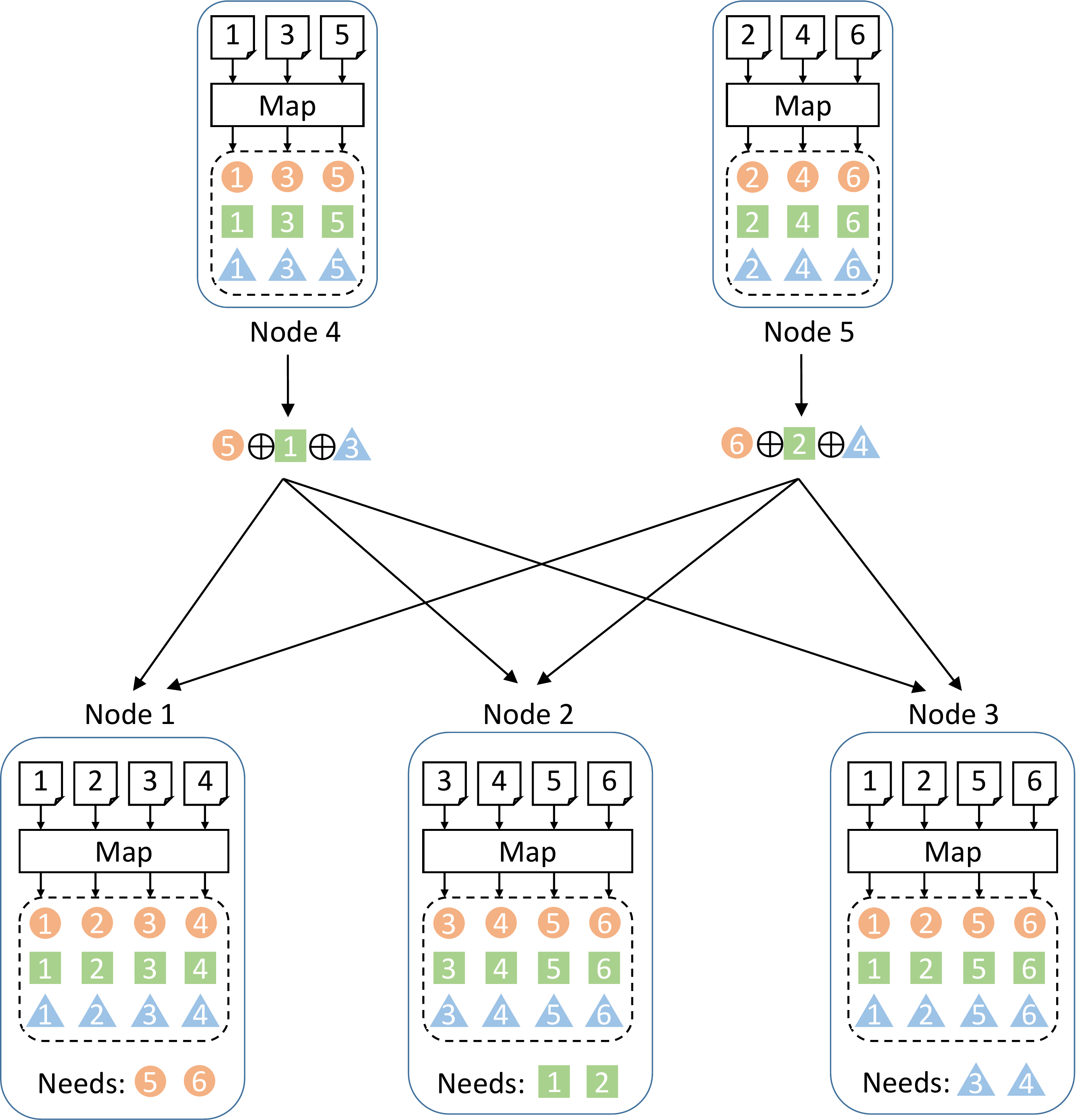}
	\caption{Illustration of the optimal achievability scheme for $N=6$, $Q=3$, $c_{\textup{m}}=1$, $c_{\textup{s}}=2$ and $c_{\textup{r}}=1$.}
	%\vspace{-3mm}
	\label{fig:users}
\end{figure}

%We present a unified scheme for a computing task with $N=6$ files, $Q=3$ Reduce functions, and parameters $K=5$, $r=2$. We use $5$ nodes, and 

\noindent {\bf Map Phase Design.} 
We let the Map task assignment to the $5$ users be $\mathcal{M}_1=\{1,2,3,4\}$, $\mathcal{M}_2=\{3,4,5,6\}$, $\mathcal{M}_3=\{1,2,5,6\}$, $\mathcal{M}_4=\{1,3,5\}$, and $\mathcal{M}_5=\{2,4,6\}$. Here each solver, i.e. users in $\{1,2,3\}$, maps $\frac{4}{6}=\frac{r^*}{Q}$
fraction of the file, and each helper maps $\frac{3}{6}<\frac{r^*}{Q}$
fraction of the files. Hence the peak computation load equals $\frac{4}{6}=\frac{r^*}{Q}$.

\noindent {\bf Shuffle Phase Design.}
After the map phase, user $4$ multicast the message $v_{5,1}\oplus v_{1,2}\oplus v_{3,3}$, and user $5$ multicast the message $v_{6,1}\oplus v_{2,2}\oplus v_{4,3}$.\footnote{Note that if network-layer multicast is not possible for delivering the coded packets, we can instead use the
existing application-layer multicast algorithms (e.g., the Message Passing Interface (MPI)) to mutlicast them (see~\cite{2016arXiv160407086L} Section VII-A for more details).} The normalized communication load equals $\frac{2}{18}=\tfrac{Q-r^*}{Q(r^*+1)}$. 
Since node $1$ knows $v_{1,2}$ and $v_{3,3}$, he can decode $v_{5,1}$ from the message multicasted by user $4$. Similarly, he can also decode $v_{6,1}$ from the other message. Because $v_{1,1},...,v_{4,1}$ are already locally computed by user $1$, the Reduce function $1$ can be executed after the shuffle phase. Same argument holds for the other $2$ Reduce functions, hence the computation can be completed after the shuffling. 

Note that in the above example, each server computes at most $1$ Reduce function. Hence the reduce time equals $1$. Consequently, the total execution time for sequential implementation equals $1\cdot\frac{4}{6}+2\cdot\frac{2}{18}+1=c_{\textup{m}}\frac{r^*}{Q}+c_{\textup{s}}\frac{Q-r^*}{Q(r^*+1)}+c_{\textup{r}}$, which can be verified to be equal to the minimum execution time $T^*_{\textup{sequential}}$ given in Theorem \ref{th:seq}.

\subsection{General Description for Sequential Implementation}

We consider a general computing task with $Q$ Reduce functions, parameters $c_\textup{m}$, $c_\textup{s}$, $c_\textup{r}$, and sufficiently large $N$.
We first compute the design parameter $r^*$ as specified in Theorem \ref{th:seq}. Depending on the value of $r^*$, we design the achievability scheme as follows.
\newline

\subsubsection{$r^*\in\{1,...,Q-1\}$} We use $K=K^*_{\textup{sequential}}$ servers as suggested in Theorem $\ref{th:seq}$. Note that $K^*_{\textup{sequential}}\geq Q$ always holds, we let nodes $1,2,...,Q$ reduce functions $1,2,...,Q$ respectively. 

%Given parameter $N$, and $Q$, for each  $r\in \{0,1,...,Q-1\}$ and $K>Q$, we construct the following map phase design:

\noindent {\bf Map Phase Design.} 
Assuming $N$ is large, we evenly partition the dataset into $(K-Q)\binom{Q}{r^*}$ disjoint subsets. We bijectively map these subsets, to tuples of a subset of $r^*$ solvers and a helper. Rigorously, we map the subset of files to the following set: $\{(i,\mathcal{A}) \ |\ i\in\{Q+1,...,K\}, \mathcal{A}\subseteq \{1,...,Q\}, |\mathcal{A}|=r^* \}$. We denote the subset of files that is mapped to $(i,\mathcal{A})$ by $\mathcal{B}_{i,\mathcal{A}}$.

%We assign these subsets of files to all $(K-Q)\binom{Q}{r}$ pairs of one non-reducer node and a subset of $r$ reducers. For each non-reducer node $i$, and each subset of $r$ reducers $\mathcal{A}$, we denote the subset of files assigned to $i$ and $\mathcal{A}$ by $\mathcal{W}_{i,\mathcal{A}}$. 
We let each solver $k\in\{1,...,Q\}$ map all subsets of files $\mathcal{B}_{i,\mathcal{A}}$ satisfying $k\in\mathcal{A}$, and we let each helper $k\in\{Q+1,...,K\}$ map all subsets $\mathcal{B}_{k,\mathcal{A}}$. Each solver maps $\frac{\binom{Q-1}{r^*-1}(K-Q)}{\binom{Q}{r^*}(K-Q)}=\frac{r^*}{Q}$ fraction of the files, and each helper maps $\frac{\binom{Q}{r^*}}{\binom{Q}{r^*}(K-Q)}=\frac{1}{K-Q}\leq \frac{r^*}{Q}$ fraction of the files. Hence, the computation time of this given Map phase design equals $c_{\textup{m}}  \frac{r^*}{Q}$.

%Map load: %The per subfile computational complexity of the above scheme is $r$ because each reducer maps $Nr$ subfiles and each additional server maps $N/\lceil1/r\rceil$ subfiles.
\noindent {\bf Shuffle Phase Design.}
	We group all the intermediate values for a Reduce function $q$ from all files in $\mathcal{B}_{i,\mathcal{A}}$ into a single variable, and denote it by $V_{i,\mathcal{A},q}$.
	At the shuffling phase, each helper from $i\in\{Q+1,...,K\}$ will multicast the following messages:
	For each subset of $r^*+1$ solvers, denoted by $\mathcal{S}$, %and for each $k\in\mathcal{S}$, 
	helper $i$ multicasts $Y_{i,\mathcal{S}}\triangleq\oplus_{k\in\mathcal{S}} {V_{i,\mathcal{S}\backslash \{k\},k}}$ to all the solvers in $\mathcal{S}$. The normalized communication load equals $\frac{\binom{Q}{r^*+1}(K-Q)}{\binom{Q}{r^*}(K-Q)Q}=\frac{Q-r^*}{Q(r^*+1)}$. Hence, the computation time of this given Shuffle phase design equals $c_{\textup{s}}  \frac{Q-r^*}{Q(r^*+1)}$.
	
	Now we prove the validity of the above scheme: For each subset $\mathcal{A}\subseteq \{2,...,Q\}$ of size $r^*$ and for each $i\in\{Q+1,...,K\}$, server $1$ can decode $V_{i,\mathcal{A},1}$ from $Y_{i,\mathcal{S}\cup\{1\}}$. Combining with the intermediate values that are computed locally on server $1$, the reduce function $1$ can be executed after the shuffle phase. Same argument holds for the other $Q-1$ Reduce functions, hence the proposed shuffling scheme is valid. 
	%Based on the broadcasts, each reducers $k$ can decode $V_{i,\mathcal{B}\backslash \{k\},k}$. (decodaility)
	
	%Hence, after receiving and decoding all the broadcasts, all reducers are able to execute the reduce function, so the above map-reduce scheme is valid.
	
	%Because the total number of packets being transmitted equals $\lceil1/r\rceil \binom{Q}{Qr} \frac{Q(1-r)}{Qr+1}$, the per subfile communication load of the above scheme equals $f(Q,r)$
	
	\begin{remark} Note that if we view all the helpers as $1$ super node, the node maps all the files and broadcasts all messages during the shuffle phase. By viewing the super node as the server and the solvers as the users, we recover the caching scheme proposed in \cite{maddah-ali12a}. In our proposed distributed computing scheme, we split the work in the map phase for the super node onto multiple nodes, in order to ensure the peak computation load is not bottlenecked by the Map tasks executed at these helpers.
	%	the message is the same, but here we have enc req. Except we dived the required computation for broadcasting server properly, st no server has to compute a lot, yet all coding gain can be obtained.
	\end{remark}

\subsubsection{$r^*=0$} In this case, Theorem \ref{th:seq} states that $T^*_{\textup{sequential}}$ cannot be exactly achieved using finite number of servers. Hence we consider picking a parameter $K$ as large as possible, and use $K$ servers for the achievability scheme. We let nodes $1,2,...,Q$ reduce functions $1,2,...,Q$ respectively, and not being assigned any Map tasks. Assuming $N$ is large, we evenly partition the dataset into $K-Q$ subsets of files, and we let each helper disjointly maps one subset. The peak computation load consequently equals $\frac{1}{K-Q}$, which is negligible if $K$ is sufficiently large. Hence the Map Phase design requires a computation time of $c_{\textup{m}}\cdot 0= c_{\textup{m}}\cdot\frac{r^*}{Q}$. 

At the shuffling phase, note that each the intermediate value is computed by exactly one helper, we simply let all the helpers unicast each intermediate value to the solver that requires the value to execute the reduce function. Because each intermediate value is unicast exactly once, the normalized communication load equals $1$ and the communication time equals $c_{\textup{s}}\cdot 1 =c_{\textup{s}}  \frac{Q-r^*}{Q(r^*+1)}$.

\subsubsection{$r^*=Q$} In this case, $K^*_{\textup{sequential}}=Q$. We simply use $Q$ servers, each reducing one function, and maps the entire database. The peak computation load equals $1$, hence the computation time equals $c_{\textup{m}}\cdot 1= c_{\textup{m}}\cdot\frac{r^*}{Q}$. Note that each server obtains all the needed intermediate values after the Map phase, no communication is required in the shuffling phase. Hence the communication time equals $c_{\textup{s}}\cdot 0 =c_{\textup{s}}  \frac{Q-r^*}{Q(r^*+1)}$. 

~\newline

In all the above cases, each server reduces at most one function. Hence our proposed achievability scheme always achieve a reduce time of $c_\textup{r}$. Besides, in all the cases, the achievability scheme uses $K^*_{\textup{sequential}}$ servers (or sufficiently many servers if  $K^*_{\textup{sequential}}$ does not exist), achieves a computation time of $ c_{\textup{m}}\cdot\frac{r^*}{Q}$ and a communication time of $c_{\textup{s}}  \frac{Q-r^*}{Q(r^*+1)}$. The total execution time always equals $T^*_{\textup{sequential}}=c_{\textup{m}}\cdot\frac{r^*}{Q}+c_{\textup{s}}  \frac{Q-r^*}{Q(r^*+1)}+c_\textup{r}$. Hence, our proposed scheme always achieves the $T^*_{\textup{sequential}}$ and $K^*_{\textup{sequential}}$ stated in Theorem \ref{th:seq}.

%and parameters $K>Q$, $r\in\{0,1,...,Q-1\}$, we present a general unified scheme using $K$ nodes, that satisfy all requirements stated at the beginning of the section.  %In the following distributed computing scheme, we always assume each user in $\{1,...,Q\}$ reduces $1$ key, and user $i\in \{1,...,Q\}$ reduces key $i$.

	\begin{remark} 	
		Interestingly, in the proposed optimal computing scheme, the minimum cost is achieved by completely separating the Reduce tasks and the shuffle jobs onto different servers. Because no solver in the proposed scheme are responsible for multicasting messages in the delivery phase, the Map tasks on the solvers can be perfectly designed in order to fully exploiting the multicast opportunity, without having to considerate the encodability constraint.  
	\end{remark}

\section{Converse}\label{sec:conv}
In this section, we derive matching converses that shows the optimality of the proposed computation scheme. We also show that our proposed optimal scheme uses the minimum possible number of nodes to achieve the minimum execution time.% can be achieved using a finite number of nodes, and what is the minimum number of required node our proposed optimal scheme uses the minimum number of nodes.

\subsection{Key Lemma}
%TODO: remarks for general lemma: 1. simple form, can be generlized to other delivery problems [][], special forms gives matching bound. 2.  works for non-bits.
%TODO: s is non-empty, d can be empty

Before deriving the exact converse for each implementation, we first prove the following key lemma, that applies for both sequential and parallel implementations. The lemma lower bounds the shuffling time given an arbitrary Map and Reduce task allocation:
\begin{lemma}[Converse Bound for Communication Load]
	\label{keyl}
	Consider a distributed computing task with $N$ files and $Q$ Reduce functions, and a given map and reduce design that uses $K$ nodes. For any integers $s,d$, let $a_{s,d}$ denotes the number of intermediate values that are available at $s$ nodes, and required by (but not available at) $d$ nodes. The following lower bound on the communication load holds:
	\begin{align}
	L\geq \frac{1}{QN}\sum_{s=1}^{K}\sum_{d=1}^{K-s} a_{s,d} \frac{d}{s+d-1}
	\end{align}
\end{lemma}

\begin{remark}
Prior to this work, several bounding techniques have been proposed for coded distributed computing and coded caching with uncoded prefetching \cite{2016arXiv160407086L, globedcd16, li2016scalable, kai2016optimality, wan2016caching, yu2016exact} . All of them can be derived as special cases of the above simple lemma.
\end{remark}

\begin{remark} Although we assume that each server sends messages independently during the shuffling phase, the above lemma can be easily generalized to computing models where the data shuffling process can be carried out in multiple rounds and dependency between
messages are allowed. We can prove that even multiple round communication is allowed, the exactly same lower bound stated in Lemma \ref{keyl} still holds. Consequently, requiring the servers communicating independently does not induce any cost in the total execution time.
\end{remark}

We postpone the proof of Lemma \ref{keyl} to Appendix \ref{app:lm1}, and in this section, we assume the correctness of this lemma and prove the optimality of the proposed schemes based on that.
%Besides, we also prove a simple lemma that uniquely determines the optimal reduce job allocation:
%\begin{lemma}[Optimal Key Assignment]
%Lemma: for sequential implementation or parallel implementation, any optimal scheme satisfies that each node reduces at most one key
%	Consider a distributed computing task with $N$ files, $Q$ keys. 
%\end{lemma}

\subsection{Converse Bounds for Sequential Implementation}

Now we use Lemma \ref{keyl} to prove a matching converse for Theorem \ref{th:seq}, which is equivalent to prove the following two statements:

\begin{enumerate}
    \item  The execution time of any coded computing scheme for a distributed computing task with $N$ files and $Q$ Reduce functions with sequential implementation is at least $T^*_{\textup{sequential}}$.  
    \item Any computing scheme that arbitrarily closely achieve a execution time of $T^*_{\textup{sequential}}$ uses at least $K^*_{\textup{sequential}}$ servers. 
\end{enumerate}

First of all, note that for any coded computing scheme, we can construct an alternative valid scheme with the same computation load and communication load, but each server only reduces at most $1$ function. The construction is given as follows:

Given the computing scheme, for each server $k$ that reduce at least $1$ functions, let $q_k$ denotes the number of functions reduced by this server. Make $q_k-1$ extra copies of this server mapping the same set of files, but not responsible for any shuffling job, and let each of these $q_k$ users reduce only one of the $q_k$ functions originally assigned to server $k$. If all map, shuffle, and reduce phases for the other servers remain the same, each additional server can still obtain enough information to execute the reduce function. Besides, the Map time and the Shuffle time remain the same, but each server in the new computing scheme only reduces at most $1$ function.

Consequently, for any computing scheme that assigns more than $1$ function to any single server, we can find a further optimized scheme with a strict improvement in the execution time of at least $c_{\textup{r}}$. Hence any such scheme can not achieve the minimum possible execution time.
So to prove a matching converse for Theorem \ref{th:seq}, it is sufficient to focus on computing schemes where each server reduces at most one function. 

We consider an arbitrary computing scheme that maps $N$ files, uses $K$ servers and reduces $Q$ functions. Without loss of generality, we assume servers in $\{1,...,Q\}$ are assigned Reduce tasks.

We first derive a lowerbound on the communication load by enhancing the computing system: We view the servers in $Q+1,...,K$ as a super node, that maps all files that are mapped by these servers, and broadcast all messages that are broadcast by these servers during the shuffling phase.\footnote{If $K=Q$, we simply let the super node not being assigned any tasks.} It is easy to verify that by enhancing the computing system in this way, all solvers are still able to execute the reduce function, and the total communication load does not increase.

We then apply Lemma \ref{keyl} on the enhanced computing system. Let $a_{j,0}$ denotes the number of files that are mapped by $j$ solvers, but not mapped by the super node, and let $a_{j,1}$ be the number of files that are mapped by $j$ solvers, and mapped by the super node. From Lemma \ref{keyl}, the communication load is lower bounded by the following inequality:
\begin{align}
L\geq \frac{1}{QN}\sum_{j=0}^{Q} (Q-j)a_{j,0}\frac{1}{j}+(Q-j)a_{j,1}\frac{1}{j+1}.
\end{align}

Note that the peak computation load is lower bounded by the average computation load on the solvers, thus

\begin{align}
p\geq \sum_{k=1}^Q\frac{|\mathcal{M}_k|}{QN}=\frac{1}{QN}\sum_{j=0}^{Q} j(a_{j,0}+a_{j,1}).\label{comp:seq}
\end{align}

Hence, the total execution time is lower bounded by 
\begin{align}
T_{\textup{sequential}}\geq &\frac{1}{QN}(\sum_{j=0}^{Q} a_{j,0}(c_\textup{m}j+c_{\textup{s}} \frac{Q-j}{j})+a_{j,1}(c_\textup{m}j+c_{\textup{s}}\frac{Q-j}{j+1}))+c_\textup{r}.
\end{align}

Note that $a_{j,0}$, $a_{j,1}$ are non-negative and satisfy the following equation
\begin{align}
N=\sum_{j=0}^{Q} (a_{j,0}+a_{j,1}).
\end{align}

Consequently, the minimum value that $T_{\textup{sequential}}$ can take is given by
\begin{align}
T_{\textup{sequential}}\geq &\frac{1}{Q}(\min_{j\in\{0,...,Q\}} \min\{ c_\textup{m}j+c_{\textup{s}} \frac{Q-j}{j}, c_\textup{m}j+c_{\textup{s}}\frac{Q-j}{j+1}\})+c_\textup{r}\\
=& \min_{r\in\{0,...,Q\}} (c_\textup{m}\frac{r}{Q}+c_{\textup{s}} \frac{Q-r}{Q(r+1)})+c_\textup{r}\\
=& T^*_{\textup{sequential}},
\end{align}
which proves the first statement.

Let $\mathcal{R}^*=\underset{r\in\{0,1,...,Q\}}{\textup{argmin}} (c_\textup{m}\frac{r}{Q}+c_{\textup{s}} \frac{Q-r}{Q(r+1)})$, we have $r^*=\max \mathcal{R}^*$.
If $T^*_{\textup{sequential}}$ is arbitrarily closely achieved, the Map task assignment of the computation scheme must satisfy that $a_{j,i}\approx 0$ except for $j\in \mathcal{R}^*$, and $i=1$ if $j\neq Q$.

We consider the following two possible cases, distinguished by the value of $r^*$:

1. If $r^*\neq Q$, i.e., $Q\notin\mathcal{R}^*$. $a_{j,i}$ can only be non-zero when $i=1$, which means almost all files must be mapped at the super node. Since the equality for (\ref{comp:seq}) must hold in order for a computing scheme to arbitrarily achieve the lower bound of $T_{\textup{sequential}}$, the peak computation load must be no larger than $\frac{r*}{Q}$. Consequently, the minimum number of helpers must be at least $\lceil\frac{1}{p}\rceil=\lceil\frac{Q}{r^*}\rceil$ in order for them to map all the files. 

Hence, we have 
\begin{align}
K\geq Q+\lceil\frac{Q}{r^*}\rceil=K^*_{\textup{sequential}}.
\end{align}
Note that if $r^*=0$, the minimum execution time can not be achieved using finite number of servers.

2. If $r^*= Q$, the required number of servers to achieve $T^*_{\textup{sequential}}$ is simply bounded by $Q$, because $Q$ Reduce functions has to be assigned to distinct servers. Hence $K\geq Q=K^*_{\textup{sequential}}$.

Hence, the second statement is proved for all possible values of $r^*$.

\subsection{Converse Bounds for Parallel Implementation}

Now we use Lemma \ref{keyl} to prove a matching for Theorem \ref{th:par}, which is equivalent to prove the following two statements:

\begin{enumerate}
    \item  The execution time of any coded computing scheme for a distributed computing task with $N$ files and $Q$ Reduce functions with parallel implementation is at least $T^*_{\textup{parallel}}$.  
    \item Any computing scheme that arbitrarily closely achieve a execution time of $T^*_{\textup{parallel}}$ uses at least $K^*_{\textup{parallel}}$ servers.
\end{enumerate}

%1. The execution time of any coded computing scheme for a distributed computing task with $N$ files and $Q$ Reduce functions with parallel implementation is at least $T^*_{\textup{parallel}}$.  

%2. Any computing scheme that arbitrarily closely achieve a execution time of $T^*_{\textup{parallel}}$ uses at least $K^*_{\textup{parallel}}$ servers. 

Similar to the sequential case, we can easily show that any computing scheme that assigns more than $1$ Reduce function to any single server can not achieve the minimum possible execution time. 
So to prove a matching converse, it is sufficient to focus on computing schemes where each server reduces at most one function. 

We consider an arbitrary a computing scheme that maps $N$ files, uses $K$ servers and reduces $Q$ functions. Without loss of generality, we assume servers in $\{1,...,Q\}$ are assigned Reduce tasks.
Following the same arguments and the same notation used for the sequential case, the following bounds for the communication load and the computation load also hold for sequential implementation:
\begin{align}
L&\geq \frac{1}{QN}\sum_{j=0}^{Q} (Q-j)a_{j,0}\frac{1}{j}+(Q-j)a_{j,1}\frac{1}{j+1},\\
p&\geq \frac{1}{QN}\sum_{j=0}^{Q} j(a_{j,0}+a_{j,1}).
\end{align}

Let $\textup{Conv}(f(\cdot))$ denotes the lower convex envelop of points $\{(r,f(r))\ |\ r\in\{0,1,...,Q\}\}$, we have
\begin{align}
L&\geq \frac{1}{N}\sum_{j=0}^{Q} (a_{j,0}+a_{j,1})\frac{Q-j}{Q(j+1)}\\
&=\frac{1}{N}\sum_{j=0}^{Q} (a_{j,0}+a_{j,1})\ \textup{Conv}\left(\frac{Q-j}{Q(j+1)}\right).
\end{align}

Note that 
\begin{align}
N=\sum_{j=0}^{Q} (a_{j,0}+a_{j,1}),
\end{align}
and $\frac{Q-j}{Q(j+1)}$ is a decreasing sequence, using Jensen's inequality, we have
\begin{align}
L&\geq \textup{Conv}\left(\frac{Q-r}{Q(r+1)}\right),\label{ineq:Jensen}
\end{align}
where $r=Qp$.

Consequently, 
\begin{align}
T_{\textup{parallel}}&\geq \min_{r\in[0,Q]} \max\{c_{\textup{m}}\frac{r}{Q}, \ c_{\textup{s}}\  \textup{Conv}\left(\frac{Q-r}{Q(r+1)}\right)\} +c_{\textup{r}}\\
&=T^*_{\textup{parallel}},
\end{align}
which proves the first statement.

It is easy to show that the above bound is minimized by a unique value $r^*\in(0,Q)$. 
If $T^*_{\textup{parallel}}$ is arbitrarily closely achieved, the equality of the Jensen's inequality used in (\ref{ineq:Jensen}) must hold. Consequently, the Map task assignment of the computation scheme must satisfy that $a_{j,i}\approx 0$ except for $j= \lfloor r^*\rfloor$ or $\lceil r^*\rceil$, and $i=1$ if $j\neq Q$.

We consider the following two possible cases, distinguished by the value of $r^*$:

1. If $r^*\leq Q-1$, $a_{j,i}$ can only be non-zero when $i=1$, which means almost all files must be mapped at the super node. Similar to the sequential case, the minimum number of helpers must be at least $\lceil\frac{1}{p}\rceil=\lceil\frac{Q}{r^*}\rceil$ in order for them to map all the files. 
Hence, we have 
\begin{align}
K\geq Q+\lceil\frac{Q}{r^*}\rceil=K^*_{\textup{parallel}}.
\end{align}

2. If $r^*> Q-1$, only $a_{Q-1,1}$, $a_{Q,0}$ and $a_{Q,1}$ can be non-zero. Hence we have 
\begin{align}
a_{Q-1,1}+a_{Q,0}+a_{Q,1}&=N\\
(Q-1)a_{Q-1,1}+Q\ a_{Q,0}+Q\ a_{Q,1}&=r^*N
\end{align}

Note that $a_{Q-1,1}+ a_{Q,1}$ files are mapped at the super node, 
the required number of servers to achieve $T^*_{\textup{sequential}}$ can be bounded as follows:
\begin{align}
K&\geq Q+\lceil\frac{a_{Q-1,1}+ a_{Q,1}}{r^*N/Q}\rceil \\
&\geq Q+\lceil\frac{a_{Q-1,1}}{r^*N/Q}\rceil\\
&=Q+\lceil\frac{QN-r^*N}{r^*N/Q}\rceil\\
&=Q+\lceil\frac{Q(Q-r^*)}{r^*}\rceil\\
&=K^*_{\textup{parallel}}.
\end{align}

Hence, the second statement is proved for all possible values of $r^*$.

%TODO: N to E, K to \tilde{K}
%\begin{remark}
%	The settings of Lemma \ref{keyl} form an abstraction of a communication system with $\tilde{K}$ nodes, sharing an error-free boradcast link. $T$ independent subfiles exist in this system, and each subfile $i$ is stored by nodes in $\mathcal{S}_i$ and requested by nodes in $\mathcal{D}_i$. All $\tilde{K}$ nodes can independently construct messages based on their local memories in order to complete the requests. In that sense the above lemma provides a lower bound on the communication load for any valid scheme, %when no feedback is used during the comminucation process
%	where messages broadcast by different nodes are constructed independently. Using the same proving technique, one can also show that allowing multiple stages of communication (and hence dependencies among messages in different stages) does not help reduce the communication load. %(also randomized encoding)
%\end{remark}

%\begin{remark}
%	Lemma \ref{keyl} also gives tight lower bounds for problems proposed in \cite{DBLP:journals/corr/LiMA15}[][], closing the constant gap between the upper bounds and the lower bounds (detailed proof included in \cite{2016arXiv160407086L}). [globecom] also uses a special form of this lemma to derive tight converse bounds.
%\end{remark}

\section{Conclusion and Future Directions}\label{sec:conc}

In this paper, we considered the problem of optimally allocating computing resources for distributed computation tasks. We proposed the optimal resource allocation scheme that minimizes the total execution time of the computation tasks, and proved its optimality through information-theoretic converses. Similarly, we proved that our proposed design uses the minimum possible number of servers among all possible computation schemes that achieves the minimum execution time. 

This work leads to several interesting future directions. From a practical perspective, we can apply and implement our proposed scheme to many distributed computing algorithms to improve their performances. One example being the TeraSort algorithm, of which the coded version has been successfully implemented \cite{hadoopsort, li2017coded}. On the other hand, we can extend this problem to a heterogeneous setting, where the processing speeds of the computing nodes varies significantly. For example, an interesting problem could be how to optimally allocate the computing resources for a cluster with a few ``super computers'', and abundant number of ``slower processors''. Prior to this work,  \cite{reisizadehmobarakeh2017coded} considered a distributed matrix multiplication problem, and shown that designing a computing scheme without fully exploiting the heterogeneity could significantly increase the computation latency.

%\begin{itemize}
%\item {\emph{Software Implementation}} Given the optimal computing scheme we proposed for the resource allocation problem, we can apply this design to many distributed computing algorithms to improve their performances. One example being the TeraSort algorithm, of which the coded version has been successfully implemented \cite{hadoopsort, li2017coded}. 
%\item {\emph{Heterogeneous Computing Environments}} In real world computing clusters, it is common that the processing speeds of the computing nodes varies significantly, and allocating the computation resources without fully exploiting the heterogeneity could be strictly suboptimal. Hence, finding the optimal resource allocation scheme in the heterogeneous setting would be an interesting open problem (e.g., how to optimally allocate the computing resources for a cluster with a few ``super computers'', and abundant number of ``slower processors''). 
%\item {\emph{Computing in General Network Topology}} In many  , the processors are setting, e.g. fattree or jellyfish
%\end{itemize}

\section{acknowledgement}
This work is in part supported by NSF grants CAREER 1408639 and NETS-1419632, ONR award N000141612189, NSA award, and funds from Intel.

\appendices

\section{Achievability schemes for the parallel implementation}\label{app:par}

		In this appendix, we provide achievability schemes that achieves the minimum execution time $T^*_{\textup{parallel}}$ for parallel implementation using $K^*_{\textup{parallel}}$ servers.
    We consider a general computing task with $Q$ Reduce functions, parameters $c_\textup{m}$, $c_\textup{s}$, $c_\textup{r}$, and sufficiently large $N$.
    We compute the design parameters $r^*$ and $K^*_{\textup{parallel}}$ specified in Theorem \ref{th:par}. It is easy to show that $r^*>0$ from (\ref{rstarp}) given that $c_\textup{s}>0$, hence $K=K^*_{\textup{parallel}}$ is always well defined.
    
    We use $K=K^*_{\textup{parallel}}$ servers, as suggested in Theorem \ref{th:par}. Note that $K^*_{\textup{parallel}}\geq Q$ always holds, we let nodes $1,2,...,Q$ reduce functions $1,2,...,Q$ respectively. Depending on the value of $r^*$, we design the map phase and reduce phase as follows.
\newline

\subsubsection{$r^*\in (0,Q-1]$} For a given parameter $r^*$, we let $r_+ \triangleq\lceil r^*\rceil$, $r_-=r_+-1$ and $\alpha=r-r_-$. It is to verify that $r_+,r_-\in\{0,1,...,Q-1\}$ and $\alpha\in[0,1]$. Assuming $N$ is large, we break the dataset into two subsets, one with $\alpha N$ files, the other with $(1-\alpha)N$ files. We construct the map and shuffle phase as follows:

\noindent {\bf Map Phase Design.} 
We first consider the map task assignment for the subset of $\alpha N$ files:
We evenly partition the set of $\alpha N$ files into $(K-Q)\binom{Q}{r_+}$ disjoint subsets. We bijectively map these subsets, to tuples of a subset of $r_+$ solvers and a helper. Rigorously, we map the subset of files to the following set: $\{(i,\mathcal{A}) \ |\ i\in\{Q+1,...,K\}, \mathcal{A}\subseteq \{1,...,Q\}, |\mathcal{A}|=r_+ \}$. We denote the subset of files that is mapped to $(i,\mathcal{A})$ by $\mathcal{B}_{i,\mathcal{A}}$.

%We assign these subsets of files to all $(K-Q)\binom{Q}{r}$ pairs of one non-reducer node and a subset of $r$ reducers. For each non-reducer node $i$, and each subset of $r$ reducers $\mathcal{A}$, we denote the subset of files assigned to $i$ and $\mathcal{A}$ by $\mathcal{W}_{i,\mathcal{A}}$. 
We let each solver $k\in\{1,...,Q\}$ map all subsets of files $\mathcal{B}_{i,\mathcal{A}}$ satisfying $k\in\mathcal{A}$, and we let each helper $k\in\{Q+1,...,K\}$ map all subsets $\mathcal{B}_{k,\mathcal{A}}$. Each solver maps $\alpha \frac{\binom{Q-1}{r_+-1}(K-Q)}{\binom{Q}{r_+}(K-Q)}=\alpha \frac{r_+}{Q}$ fraction of the files, and each helper maps $\alpha \frac{\binom{Q}{r_+}}{\binom{Q}{r_+}(K-Q)}=\alpha \frac{1}{K-Q}$ fraction of the files. %Hence, the computation time of this given Map phase design equals $c_{\textup{m}}  \frac{r^*}{Q}$.

We map the rest of the $(1-\alpha)N$ files in a similar way, except we let each file be repetitively mapped by $r_-$ solvers. This requires extra computation loads of $(1-\alpha)\frac{r_-}{Q}$ on each solver and $(1-\alpha)\frac{1}{K-Q}$ on each helper. Hence, the each solver maps $\alpha \frac{r_+}{Q}+(1-\alpha)\frac{r_-}{Q}=\frac{r^*}{Q}$ fraction of the files, and each helper maps $\alpha\frac{1}{K-Q}+(1-\alpha)\frac{1}{K-Q}=\frac{1}{K-Q}\leq \frac{r^*}{Q}$ fraction of the files. The peak computation load thus equals $\frac{r^*}{Q}$ and the computation time equals $c_\textup{m} \frac{r^*}{Q}$.

\noindent {\bf Shuffle Phase Design.}
	We first consider a shuffling scheme that delivers all intermediate values computed from the subset of $\alpha N$ files:
	We group all the intermediate values for a Reduce function $q$ from all files in $\mathcal{B}_{i,\mathcal{A}}$ into a single variable, and denote it by $V_{i,\mathcal{A},q}$.
	At the shuffling phase, each helper from $i\in\{Q+1,...,K\}$ will multicast the following messages:
	For each subset of $r_++1$ solvers, denoted by $\mathcal{S}$, %and for each $k\in\mathcal{S}$, 
	helper $i$ multicasts $Y_{i,\mathcal{S}}\triangleq\oplus_{k\in\mathcal{S}} {V_{i,\mathcal{S}\backslash \{k\},k}}$ to all the solvers in $\mathcal{S}$. The normalized communication load equals $\alpha\frac{\binom{Q}{r_++1}(K-Q)}{\binom{Q}{r_+}(K-Q)Q}=\alpha\frac{Q-r_+}{Q(r_++1)}$. %Hence, the computation time of this given Shuffle phase design equals $c_{\textup{s}}  \frac{Q-r_+}{Q(r_++1)}$.
	
	The validity of the above scheme is proved as follows: For each subset $\mathcal{A}\subseteq \{2,...,Q\}$ of size $r_+$ and for each $i\in\{Q+1,...,K\}$, server $1$ can decode $V_{i,\mathcal{A},1}$ from $Y_{i,\mathcal{S}\cup\{1\}}$. Combining with the intermediate values that are computed locally, server $1$ obtained all intermediate values mapped from the files in the subset of size $\alpha N$ for reduce function $1$. %can be executed after the shuffle phase. 
	Same argument holds for the other $Q-1$ Reduce functions, hence the proposed shuffling scheme is valid for delivering the intermediate values that are mapped from the subset of $\alpha N$ files.
	
	Similarly, we can deliver the rest of the $(1-\alpha) N$ files using a communication load of $(1-\alpha)\frac{Q-r_-}{Q(r_-+1)}$. Hence the total communication time of the proposed scheme equals $c_{\textup{s}}(\alpha\frac{Q-r_+}{Q(r_++1)}+(1-\alpha)\frac{Q-r_-}{Q(r_-+1)})=c_{\textup{s}} \cdot\textup{Conv} (\frac{Q-r^*}{Q(r^*+1)})$.
	
\subsubsection{$r^*\in (Q-1,Q]$} Similar to the other case, we define parameters $r_+=Q$, $r_-=Q-1$ and $\alpha=r-r_-$, and we break the dataset into two subsets and handle the map and reduce tasks for these two subsets separately. For the subset of size $(1-\alpha) N$, we use exactly the same Map and Shuffle phase design as discussed above, which requires computation loads of $(1-\alpha) \frac{r_-}{Q}$ on each solver, $(1-\alpha) \frac{1}{K-Q}$ on each helper, and a communication load of $(1-\alpha)\frac{Q-r_-}{Q(r_-+1)}$. However for the rest of the files, we simply let all of them to be mapped on all the solvers, which requires no extra computation on the helpers and no extra communication.

The computation load on each solver thus equals $(1-\alpha) \frac{r_-}{Q}+\alpha=\frac{r^*}{Q}$, and the computation load on each helper equals $(1-\alpha) \frac{1}{K-Q}\leq \frac{r^*}{Q}$. Consequently, the computation time equals $c_\textup{m} \frac{r^*}{Q}$. On the other hand, the communication load equals, $(1-\alpha)\frac{Q-r_-}{Q(r_-+1)}=\textup{Conv} (\frac{Q-r^*}{Q(r^*+1)})$, hence the communication time equals $c_{\textup{s}} \cdot\textup{Conv} (\frac{Q-r^*}{Q(r^*+1)})$.
%Using $K$ servers for the computing task,	 the minimum execution time can be achieved by memory sharing of unified computing schemes with parameter $r=\lceil r^*\rceil $ or $\lceil r^*\rceil -1$.
		 
		%If $Q-1<r^*\leq Q$, let $K=Q+\lceil\tfrac{Q(Q-r^*)}{r^*}\rceil$, and we use $K$ servers for the computing task. We divide the dataset of files into two groups, with one group containing $(Q-r^*)$ fraction of files. We first apply the unified scheme with parameter $r=Q-1$ on the $(Q-r^*)$ fraction of files. Within this group of files, each solver maps $\frac{Q-1}{Q}$ fraction of files, and each helper maps $1/\lceil\tfrac{Q(Q-r^*)}{r^*}\rceil$ fraction of files. In addition, we let each solver maps all files in the second group, hence all intermediate values computed from those files are available for all solvers. Consequently, the computation time equals 
		%\begin{align}
		%T_\textup{map}=&\max\{(Q-r^*)\frac{Q-1}{Q}+1-Q+r^*,\frac{Q-r^*}{\lceil\tfrac{Q(Q-r^*)}{r^*}\rceil}\}\\
		%=&\frac{r^*}{Q}.
		%\end{align} 
		%Hence, the overall execution time equals $T_\textup{parallel}=\max\{c_{\textup{m}}\frac{r^*}{Q},c_{\textup{s}} \frac{Q-r^*}{Q^2}  \}+c_{\textup{r}}=T_\textup{parallel}^*$.
		~\newline
		
		In all the above cases, each server reduces at most one function. Hence our proposed achievability scheme always achieve a reduce time of $c_\textup{r}$. Besides, in all the cases, the achievability scheme uses $K^*_{\textup{sequential}}$ servers, achieves a computation time of $ c_{\textup{m}}\cdot\frac{r^*}{Q}$ and a communication time of $c_{\textup{s}} \cdot\textup{Conv} (\frac{Q-r^*}{Q(r^*+1)})$. The total execution time always equals $T^*_{\textup{sequential}}=c_{\textup{m}}\cdot\frac{r^*}{Q}+c_{\textup{s}} \cdot\textup{Conv} (\frac{Q-r^*}{Q(r^*+1)})+c_\textup{r}$. Hence, our proposed scheme always achieves the $T^*_{\textup{sequential}}$ and $K^*_{\textup{sequential}}$ stated in Theorem \ref{th:par}.

\section{Proof of Lemma \ref{keyl}}\label{app:lm1}
\begin{proof}
%	We prove that Lemma \ref{keyl} is true for any $K\in\mathbb{N}^+$, by induction:

%	\noindent \textbf{a.} If $K=1$, the following inequality holds,
%	\begin{equation}
%	L\geq 0=\sum_{s=1}^{1}\sum_{d=1}^{1-s} a_{s,d} \frac{d}{s+d-1}
%	\end{equation}
	
%	\noindent	\textbf{b.} Suppose Lemma \ref{keyl} is true when $K=K_0$, for a $K_0\in \mathbb{N}^+$. 
	
	For $q \in \{1, ..., Q\}$, $n \in \{1, ..., N\}$, we let $V_{q,n}$ be i.i.d. random variables uniformly distributed on
$\mathbb{F}_{2^T}$ . We let the intermediate values $v_{q,n}$ be the realizations of $V_{q,n}$. For any $\mathcal{Q}\subseteq\{1,...,Q\}$, and $\mathcal{N}\subseteq\{1,...,N\}$, we define
\begin{align}
V_{\mathcal{Q},\mathcal{N}}\triangleq \{V_{q,n} : q \in\mathcal{Q}, n \in \mathcal{N} \}.
\end{align}

Since each message $X_k$ is generated as a function of the intermediate values that are computed at node $k$, the following equation holds for all $k\in\{1,...,K\}$:\footnote{$[Q]\triangleq\{1,...,Q\}$.}
 	\begin{align}
 	H(X_k|V_{[Q],{\mathcal{M}_{k}}})=0.
 	\label{causality}
 	\end{align}

The validity of the shuffling scheme requires that for all $k\in\{1,...,K\}$, the following equation holds :
\begin{align}
H(V_{\mathcal{W}_k,[N]}|X_{[K]},V_{[Q],{\mathcal{M}_k}})=0.
\label{validity}
\end{align}

Given $\boldsymbol{\mathcal{M}}$ and $\boldsymbol{\mathcal{W}}$, for any disjoint subsets of users $\mathcal{S}$ and $\mathcal{D}$, we denote the number of intermediate values that are exclusively available at servers in $\mathcal{S}$, and exclusively needed by (but not available at) servers in $\mathcal{D}$, by $a_{\mathcal{S}, \mathcal{D}}$, i.e.:
\begin{align}
a_{\mathcal{S}, \mathcal{D}}=| ((\underset{k \in {\cal S}}{\cap} {\cal M}_k) \backslash (\underset{i \notin {\cal S}}{\cup} {\cal M}_i ) )\cap ( (\underset{k \in {\cal D}}{\cap} {\cal W}_k) \backslash (\underset{i \notin {\cal D}\cup {\cal S}}{\cup} {\cal W}_i ))|. 
\end{align}

For any subset $\mathcal{C}\subseteq \{1,...,K\}$, let $\mathcal{C}^\complement =\{1,...,K\} \backslash \mathcal{C}$. We define \begin{align}
Y_{\mathcal{C}^\complement} \triangleq (V_{\mathcal{W}_{\mathcal{C}^\complement},[N]},V_{[Q],\mathcal{M}_{\mathcal{C}^\complement}}).
\end{align}
We denote the number of intermediate values that are exclusively available at $s$ servers in $\mathcal{C}$, and exclusively needed by (but not available at) $d$ users in $\mathcal{C}$, by $a_{s,d,\mathcal{C}}$, i.e.:
\begin{align}
a_{s,d,\mathcal{C}}=&\sum_{\substack{\mathcal{S}\subseteq{\mathcal{C}}\\|\mathcal{S}|={s}} }\sum_{\substack{\mathcal{D}\subseteq{\mathcal{C}}\backslash \mathcal{S}\\|\mathcal{D}|={d}}} a_{\mathcal{S}, \mathcal{D}}. 
\end{align}

Then we prove the following statement by induction: 

\begin{claim}
\noindent For any subset $\mathcal{C}\subseteq \{1,...,K\}$, we have $H(X_{\mathcal{C}}|Y_{\mathcal{C}^\complement})\geq T\sum\limits_{s=1}^{|\mathcal{C}|}\sum\limits_{d=1}^{|\mathcal{C}|-s} a_{s,d,\mathcal{C}}\cdot\frac{d}{s+d-1}$.
\end{claim}

a. If $\mathcal{C} = \emptyset$, obviously
\begin{align}
H(X_\emptyset|Y_{\emptyset^c})&\geq 0	=T\sum\limits_{s=1}^{0}\sum\limits_{d=1}^{0-s} a_{s,d,\emptyset}\cdot\frac{d}{s+d-1}.
\end{align}

b. Suppose the statement is true for all subsets of size $C_0$. 

For any $\mathcal{C}\subseteq \{1,...,K\}$ of size $|\mathcal{C}|=C_0+1$, and all $k \in \mathcal{C}$, the subset version of (\ref{causality}) and (\ref{validity}) can be derived:
 	\begin{align}
 	H(X_k|V_{[Q],{\mathcal{M}_{k}}},Y_{\mathcal{C}^\complement})=0,
 	\label{scausality}
 	\end{align}
 	\begin{align}
 	H(V_{\mathcal{W}_{k},[N]}|X_\mathcal{C},V_{[Q],{\mathcal{M}_k}},Y_{\mathcal{C}^\complement})=0.
 	\label{svalidity}
 	\end{align}
 	
 	Consequently, the following equation holds:
 	\begin{align}
 	H(X_\mathcal{C}|V_{[Q],{\mathcal{M}_k}},Y_{\mathcal{C}^\complement}) =&H(X_\mathcal{C}|V_{\mathcal{W}_{k},[N]},V_{[Q],{\mathcal{M}_k}},Y_{\mathcal{C}^\complement})+H(V_{\mathcal{W}_{k},[N]}|V_{[Q],{\mathcal{M}_k}},Y_{\mathcal{C}^\complement}).
 	\label{intermediate}
 	\end{align}
 	
 	Next we lower bound $H(X_\mathcal{C}|Y_{\mathcal{C}^\complement})$ as follows:
 	\begin{align}
 		H(X_\mathcal{C}|Y_{\mathcal{C}^\complement})&=\frac{1}{|\mathcal{C}|} \sum_{k\in\mathcal{C}}H(X_\mathcal{C}, X_{k}|Y_{\mathcal{C}^\complement})\\
 		&=\frac{1}{|\mathcal{C}|} \sum_{k\in\mathcal{C}} (H(X_{\mathcal{C}}|X_k,  Y_{\mathcal{C}^\complement})+H(X_k| Y_{\mathcal{C}^\complement}))\\
 		&\geq\frac{1}{|\mathcal{C}|} \sum_{k\in\mathcal{C}} H(X_{\mathcal{C}}|X_k,  Y_{\mathcal{C}^\complement})+\frac{1}{|\mathcal{C}|}H(W_\mathcal{C}| Y_{\mathcal{C}^\complement}).\label{eq:XS}
 	\end{align}
 	
 	From (\ref{eq:XS}), we can derive a lower bound on $H(W_{\mathcal{C}}|Y_{\mathcal{C}^\complement})$ that equals the LHS of (\ref{intermediate}) scaled by $\frac{1}{C_0}$:
 	\begin{align}
 	H(X_{\mathcal{C}}|Y_{\mathcal{C}^\complement})&\geq\frac{1}{|\mathcal{C}|-1} \sum_{k\in\mathcal{C}} H(X_{\mathcal{C}}|X_k, Y_{\mathcal{C}^\complement})\\
 	&\geq\frac{1}{C_0} \sum_{k\in\mathcal{C}} H(X_{\mathcal{C}}|X_k, V_{[Q],{\mathcal{M}_k}}, Y_{\mathcal{C}^\complement})\\
 	&=\frac{1}{C_0} \sum_{k\in\mathcal{C}} H(X_{\mathcal{C}}| V_{[Q],{\mathcal{M}_k}}, Y_{\mathcal{C}^\complement}). \label{eq:all}
 	\end{align}
 	
    The first term on the RHS of (\ref{intermediate}) is lower bounded by the induction assumption:
    \begin{align}
    H(X_\mathcal{C}|V_{\mathcal{W}_k,[N]},V_{[Q],{\mathcal{M}_k}},Y_{\mathcal{S}^c})&=H(X_{\mathcal{C}\backslash \{k\}}|Y_{(\mathcal{C}\backslash \{k\})^\complement})\\    
    &\geq T\sum_{s=1}^{C_0}\sum_{d=1}^{C_0-s} a_{s,d,\mathcal{C}\backslash \{k\}} \cdot\frac{d}{s+d-1}\\
    &=T\sum_{\substack{\mathcal{S}\subseteq\mathcal{C}\backslash\{k\}\\|\mathcal{S}|\geq 1}}\sum_{\substack{\mathcal{D}\subseteq\mathcal{C}\backslash\{k\}\backslash\mathcal{S}\\|\mathcal{D}|\geq 1}} a_{\mathcal{S},\mathcal{D}} \cdot\frac{|\mathcal{D}|}{|\mathcal{S}|+|\mathcal{D}|-1}\\
    &=T\sum_{\substack{\mathcal{S}\subseteq\mathcal{C}\\|\mathcal{S}|\geq 1}}\sum_{\substack{\mathcal{D}\subseteq\mathcal{C}\backslash\mathcal{S}\\|\mathcal{D}|\geq 1}} a_{\mathcal{S},\mathcal{D}} \cdot\frac{|\mathcal{D}|\cdot \mathbbm{1}(k\notin \mathcal{S}\cup{\mathcal{D}})}{|\mathcal{S}|+|\mathcal{D}|-1}. \label{eq:first}
    \end{align} 

	The second term on the RHS of (\ref{intermediate}) can be calculated based on the independence of intermediate values:

\begin{align}
&H(V_{\mathcal{W}_k,[N]}|V_{[Q],{\mathcal{M}_k}},Y_{\mathcal{C}^\complement})\\&= H(V_{\mathcal{W}_k,[N]}|V_{[Q],{\mathcal{M}_k}},V_{\mathcal{W}_{\mathcal{C}^\complement},[N]},V_{[Q],\mathcal{M}_{\mathcal{C}^\complement}})\\
&=T\sum_{\mathcal{S}\subseteq\mathcal{C}\backslash\{k\}}\sum_{\substack{\mathcal{D}\subseteq\mathcal{C}\backslash\mathcal{S}\\k\in\mathcal{D}}} a_{\mathcal{S},\mathcal{D}} \\
&\geq T\sum_{\substack{\mathcal{S}\subseteq\mathcal{C}\backslash\{k\}\\|\mathcal{S}|\geq 1}}\sum_{\substack{\mathcal{D}\subseteq\mathcal{C}\backslash\mathcal{S}\\k\in\mathcal{D}}} a_{\mathcal{S},\mathcal{D}}\\
&= T\sum_{\substack{\mathcal{S}\subseteq\mathcal{C}\backslash\{k\}\\|\mathcal{S}|\geq 1}}\sum_{\substack{\mathcal{D}\subseteq\mathcal{C}\backslash\mathcal{S}\\|\mathcal{D}|\geq 1}} a_{\mathcal{S},\mathcal{D}} \cdot \mathbbm{1}(k\in\mathcal{D}).\label{eq:second}
\end{align}

Thus by (\ref{intermediate}), (\ref{eq:all}), (\ref{eq:first}) and (\ref{eq:second}), we have
	\begin{align}
	H(W_{\mathcal{C}}|Y_{\mathcal{C}^\complement})&\geq\frac{1}{C_0} \sum_{k\in\mathcal{C}} H(X_{\mathcal{C}}| V_{[Q],{\mathcal{M}_k}}, Y_{\mathcal{C}^\complement})\\
	&= \frac{1}{C_0} \sum_{k\in\mathcal{C}} (H(X_\mathcal{C}|V_{\mathcal{W}_{k},[N]},V_{[Q],{\mathcal{M}_k}},Y_{\mathcal{C}^\complement})+H(V_{\mathcal{W}_{k},[N]}|V_{[Q],{\mathcal{M}_k}},Y_{\mathcal{C}^\complement})) \label{eq:twoParts}\\
	&\geq \frac{T}{C_0} \sum_{k\in\mathcal{C}} \sum_{\substack{\mathcal{S}\subseteq\mathcal{C}\\|\mathcal{S}|\geq 1}}\sum_{\substack{\mathcal{D}\subseteq\mathcal{C}\backslash\mathcal{S}\\|\mathcal{D}|\geq 1}} a_{\mathcal{S},\mathcal{D}}( \frac{|\mathcal{D}|\cdot \mathbbm{1}(k\notin \mathcal{S}\cup{\mathcal{D}})}{|\mathcal{S}|+|\mathcal{D}|-1}+\mathbbm{1}(k\in\mathcal{D})) \\
	&=\frac{T}{C_0}  \sum_{\substack{\mathcal{S}\subseteq\mathcal{C}\\|\mathcal{S}|\geq 1}}\sum_{\substack{\mathcal{D}\subseteq\mathcal{C}\backslash\mathcal{S}\\|\mathcal{D}|\geq 1}} a_{\mathcal{S},\mathcal{D}} \sum_{k\in\mathcal{C}}( \frac{|\mathcal{D}|\cdot \mathbbm{1}(k\notin \mathcal{S}\cup{\mathcal{D}})}{|\mathcal{S}|+|\mathcal{D}|-1}+\mathbbm{1}(k\in\mathcal{D})) \\
	&=\frac{T}{C_0}  \sum_{\substack{\mathcal{S}\subseteq\mathcal{C}\\|\mathcal{S}|\geq 1}}\sum_{\substack{\mathcal{D}\subseteq\mathcal{C}\backslash\mathcal{S}\\|\mathcal{D}|\geq 1}} a_{\mathcal{S},\mathcal{D}} ( \frac{|\mathcal{D}|\cdot (|\mathcal{C}|-|\mathcal{S}|-|\mathcal{D}|)}{|\mathcal{S}|+|\mathcal{D}|-1}+|\mathcal{D}|) \\
	&=\frac{T}{C_0}  \sum_{\substack{\mathcal{S}\subseteq\mathcal{C}\\|\mathcal{S}|\geq 1}}\sum_{\substack{\mathcal{D}\subseteq\mathcal{C}\backslash\mathcal{S}\\|\mathcal{D}|\geq 1}} a_{\mathcal{S},\mathcal{D}} \frac{|\mathcal{D}|\cdot (|\mathcal{C}|-1)}{|\mathcal{S}|+|\mathcal{D}|-1}\\
	&=T  \sum_{\substack{\mathcal{S}\subseteq\mathcal{C}\\|\mathcal{S}|\geq 1}}\sum_{\substack{\mathcal{D}\subseteq\mathcal{C}\backslash\mathcal{S}\\|\mathcal{D}|\geq 1}} a_{\mathcal{S},\mathcal{D}} \frac{|\mathcal{D}|}{|\mathcal{S}|+|\mathcal{D}|-1}
	. \label{eq:int1}
	%&=T\sum_{j=1}^{S_0+1}  a^{j,\mathcal{S}}_{\cal M} {\frac{Q}{K}} \cdot\frac{S_0+1-j}{j}
	\end{align}

From the definition of $a_{s,d,\mathcal{C}}$ and (\ref{eq:int1}) , we have:
\begin{align}
H(W_{\mathcal{C}}|Y_{\mathcal{C}^\complement})&\geq T  \sum_{s=1}^{|\mathcal{C}|}\sum_{d=1}^{|\mathcal{C}|-s} a_{s,d,\mathcal{C}} \frac{d }{s+d-1}.
%\sum_{k\in\mathcal{S}}  a^{j,\mathcal{S}\backslash \{k\}}_{\cal U}&=\sum_{k\in\mathcal{S}} \sum_{f=1}^{N} \mathbbm{1}{(\textup{file $f$ is only stored by users in $\mathcal{S}\backslash \{k\}$})}\cdot  \mathbbm{1}{(\textup{$f$ is stored by $j$ users})}\\
%&=\sum_{f=1}^{N}  \mathbbm{1}{(\textup{file $f$ is only stored by $j$ users in ${\cal S}$})}\sum_{k\in\mathcal{S}}  \mathbbm{1}{(\textup{$f$ is not stored by User $k$})}\\
%&= \sum_{f=1}^{N}  \mathbbm{1}{(\textup{file $f$ is only stored by $j$ users in ${\cal S}$})} (|\mathcal{S}|-j)\\
%&= a^{j,\mathcal{S}}_{\cal U}(S_0+1-j). \label{eq:int2}
\end{align}
	
c. Thus for all subsets $\mathcal{C}\subseteq \{1,...,K\}$, the following equation holds:
\begin{align}
H(X_{\mathcal{C}}|Y_{\mathcal{C}^\complement})\geq T\sum_{s=1}^{|\mathcal{C}|}\sum_{d=1}^{|\mathcal{C}|-s} a_{s,d,\mathcal{C}} \frac{d }{s+d-1},
\end{align}
which proves Claim~1.

Then by Claim~1, let $\mathcal{C}=\{1,...,K\}$ be the set of all $K$ users, 
\begin{align}
L\geq \frac{H(X_{\mathcal{C}}|Y_{\mathcal{C}^\complement})}{QNT}\geq \frac{1}{QN} \sum_{s=1}^{K}\sum_{d=1}^{K-s} a_{s,d} \frac{d }{s+d-1}.
\end{align}

This completes the proof of Lemma~1.\end{proof}

%\begin{remark}
%When equality hold, $hxi=xx$
%\end{remark}

\bibliographystyle{IEEEtran}
\bibliography{ICC2017}

\end{document}